\documentclass[11pt,a4paper]{article}

\mathchardef\mhyphen="2D
\def\xsection{\ensuremath{\sigma(pp \to \bar{t}H^{+} + X)}}
\def\Hp{\ensuremath{H^{+}}}

\def\taup{\ensuremath{\tau^{+}}}

\def\taupm{\ensuremath{\tau^{\pm}}}
\def\brhtn{\ensuremath{{\cal B}(\Hp \to \taup \nu)}}
\def\mhmax{$m_h^{\mathrm{max}}$}
\def\mhmodp{$m_h^{\mathrm{mod+}}$}
\def\mhmodm{$m_h^{\mathrm{mod-}}$}

\newcommand{\LUMI}{19.5\,fb$^{-1}$ }
\def\mt{\ensuremath{m_{\mathrm{T}}}}

\def\jvfcut{\ensuremath{|}JVF\ensuremath{|>0.5}}
\def\jvfcut2{\ensuremath{\mathrm{JVF}>0.5}}

\def\tauhadvis{\ensuremath{\tau_{\mathrm{had}\mhyphen\mathrm{vis}}}}
\def\tauhad{\ensuremath{\tau_{\mathrm{had}}}}

\def\mtp2{\ensuremath{m_{\mathrm{T}2}^{H}}}

\usepackage{preprintcover}
\PreprintCoverPaperTitle{\boldmath Search for charged Higgs bosons decaying
via $H^{\pm} \rightarrow \taupm \nu$ in fully hadronic final states
using $pp$ collision data at $\sqrt{s} = 8~\mbox{TeV}$
with the ATLAS detector}  % paper title

\PreprintIdNumber{CERN-PH-EP-2014-274}  % CERN preprint number ---> can be found in CDS entry for Draft 2!

\PreprintCoverAbstract{
The results of a search for charged Higgs bosons decaying to a $\tau$ lepton
and a neutrino, $H^{\pm} \to \tau^{\pm} \nu$,  are presented.
The analysis is based on 19.5 fb$^{-1}$ of proton--proton collision data at
$\sqrt{s}=8$ TeV collected by the ATLAS experiment at the Large Hadron
Collider.
Charged Higgs bosons are searched for in events consistent
with top-quark pair production or in associated production with
a top quark, depending on the considered $H^{\pm}$ mass. The final state is characterised by the presence
of a hadronic $\tau$ decay, missing transverse momentum, $b$-tagged jets, a
hadronically decaying $W$ boson, and the absence of any isolated electrons or muons with
high transverse momenta.
The data are consistent with the expected background from Standard
Model processes.
A statistical analysis leads to 95\% confidence-level upper
limits on the product of branching ratios $ {\cal B}(t\rightarrow bH^\pm) \times  {\cal B}(H^\pm\rightarrow \tau^{\pm}\nu) $,
between 0.23\% and 1.3\% for charged Higgs boson masses in the range
80--160 GeV. It also leads to 95\% confidence-level upper limits on
the production cross section times branching ratio, $\sigma(pp \to tH^{\pm} + X) \times {\cal B}(H^{\pm} \rightarrow \tau^{\pm} \nu)$, between
0.76 pb and 4.5 fb, for charged Higgs boson masses ranging from
180 GeV to 1000 GeV. In the context of different
scenarios of the Minimal Supersymmetric Standard Model, these results exclude nearly all values of
$\tan\beta$ above one for charged Higgs boson masses between 80 GeV
and 160 GeV, and exclude a region of parameter space with high $\tan\beta$ for $H^{\pm}$ masses between 200 GeV and 250 GeV.
}

\PreprintJournalName{JHEP}

\usepackage{jheppub}

\usepackage[numbers,sort&compress]{natbib}
\usepackage{atlasphysics} 
\usepackage{lineno}
\usepackage{hyperref}
\usepackage{subfigure}
\usepackage{mathrsfs}
\usepackage{bm}
\usepackage{placeins}

\makeatletter
\g@addto@macro\bfseries{\boldmath}
\makeatother

\begin{document}

\clearpage
% Title
\title{\boldmath Search for charged Higgs bosons decaying 
via $H^{\pm} \rightarrow \tau^{\pm} \nu$ in fully hadronic final states 
using $pp$ collision data at $\sqrt{s} = 8~\mbox{TeV}$ 
with the ATLAS detector}
% Author
\author{The ATLAS Collaboration}
%draft version 7.0}

\abstract{
The results of a search for charged Higgs bosons decaying to a $\tau$ lepton 
and a neutrino, $H^{\pm} \to \tau^{\pm} \nu$,  are presented.
The analysis is based on 19.5\,fb$^{-1}$ of proton--proton collision data at
$\sqrt{s}=8$~TeV collected by the ATLAS experiment at the Large Hadron
Collider.
Charged Higgs bosons are searched for in events consistent
with top-quark pair production or in associated production with
a top quark, depending on the considered $H^{\pm}$ mass. The final state is characterised by the presence
of a hadronic $\tau$ decay, missing transverse momentum, $b$-tagged jets, a
hadronically decaying $W$ boson, and the absence of any isolated electrons or muons with 
high transverse momenta. 
The data are consistent with the expected background from Standard
Model processes. 
A statistical analysis leads to 95\% confidence-level upper
limits on the product of branching ratios $ {\cal B}(t\rightarrow bH^\pm) \times  {\cal B}(H^\pm\rightarrow \tau^{\pm}\nu) $, 
between 0.23\% and 1.3\% for charged Higgs boson masses in the range 
80--160 GeV. It also leads to 95\% confidence-level upper limits on
the production cross section times branching ratio, $\sigma(pp \to tH^{\pm} + X) \times {\cal B}(H^{\pm} \rightarrow \tau^{\pm} \nu)$, between
0.76\,pb and 4.5\,fb, for charged Higgs boson masses ranging from
180 GeV\ to 1000 GeV. 
In the context of different
scenarios of the Minimal Supersymmetric Standard Model, these results exclude nearly all values of
$\tan\beta$ above one for charged Higgs boson masses between 80 GeV\
and 160 GeV, and exclude a region of parameter space with high $\tan\beta$ for $H^{\pm}$ masses between 200 GeV\ and 250 GeV.
%In a flipped Two Higgs Doublet Model,  $H^+$ masses between 200--600\,GeV are excluded 
%for $\tan\beta=50$ and $|\cos(\beta-\alpha)|<0.3$.
}

\maketitle

%%%%%%%%%%%%%%%%%%%%%%%%%%%%%%%%%%%%%%%%%%%%%%%%%%%%%%%%%%%%%%%%%%%%%%%%%%%%%%%
% Introduction
%%%%%%%%%%%%%%%%%%%%%%%%%%%%%%%%%%%%%%%%%%%%%%%%%%%%%%%%%%%%%%%%%%%%%%%%%%%%%%%

\section{Introduction}
\label{Introduction}

\begin{figure}[h!]
\begin{center}
\subfigure[\label{fig:feynmanI}low-mass $H^+$ production]{
  \includegraphics[width=0.3\textwidth]{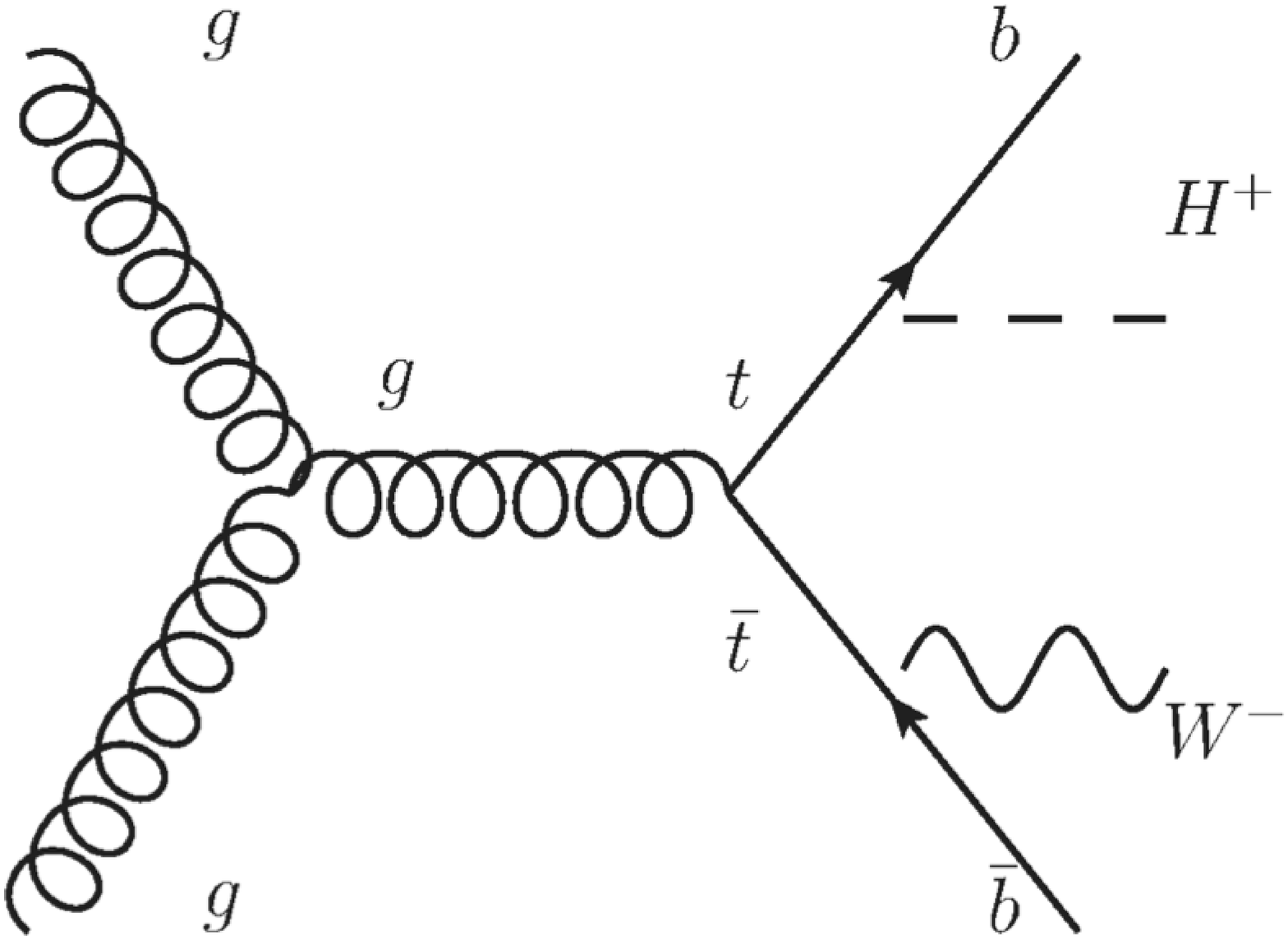}
}
\subfigure[\label{fig:feynmanI}5FS high-mass $H^+$ production]{
  \includegraphics[width=0.3\textwidth]{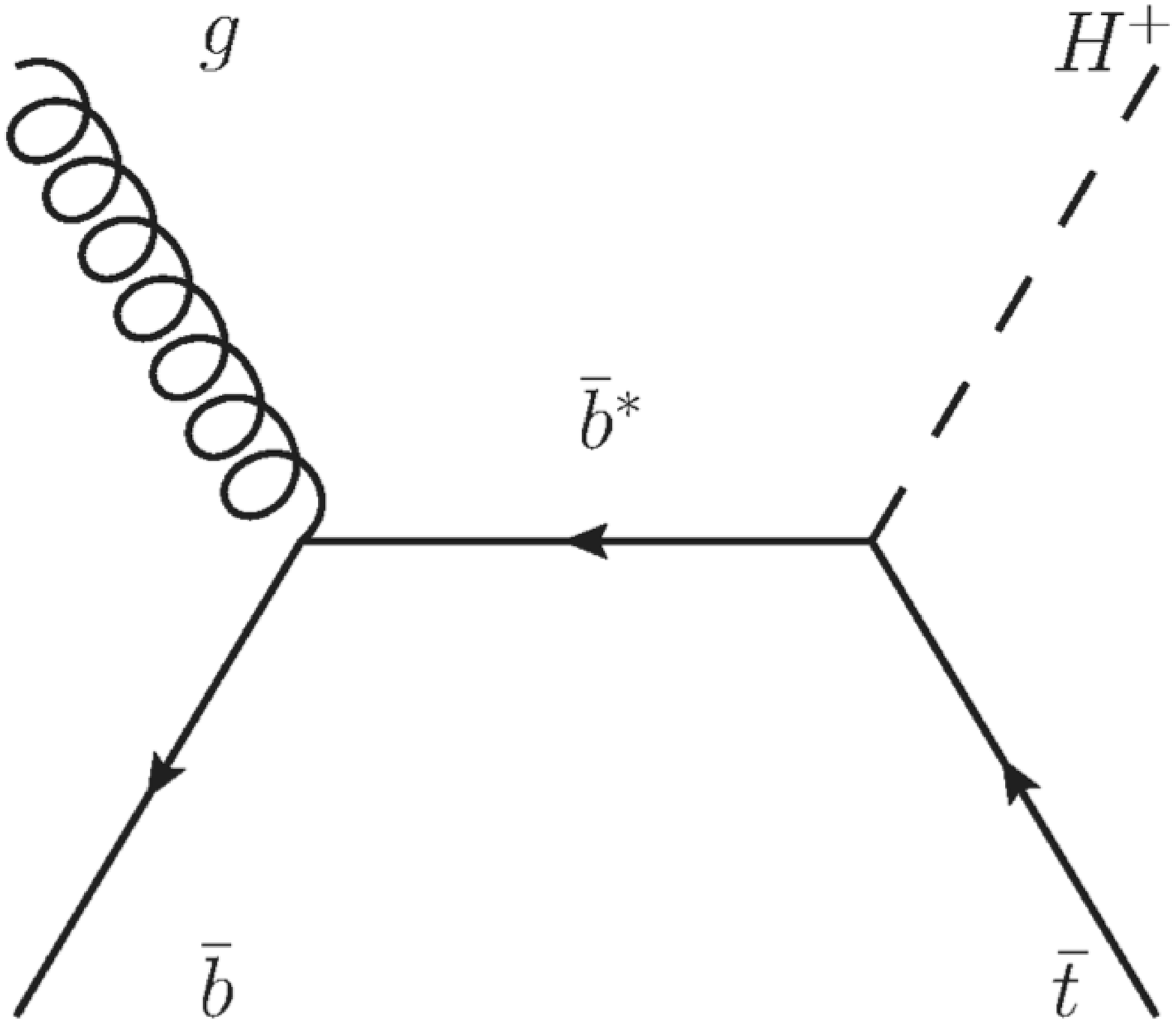}
}
\subfigure[\label{fig:feynmanI}4FS high-mass $H^+$ production]{
  \includegraphics[width=0.3\textwidth]{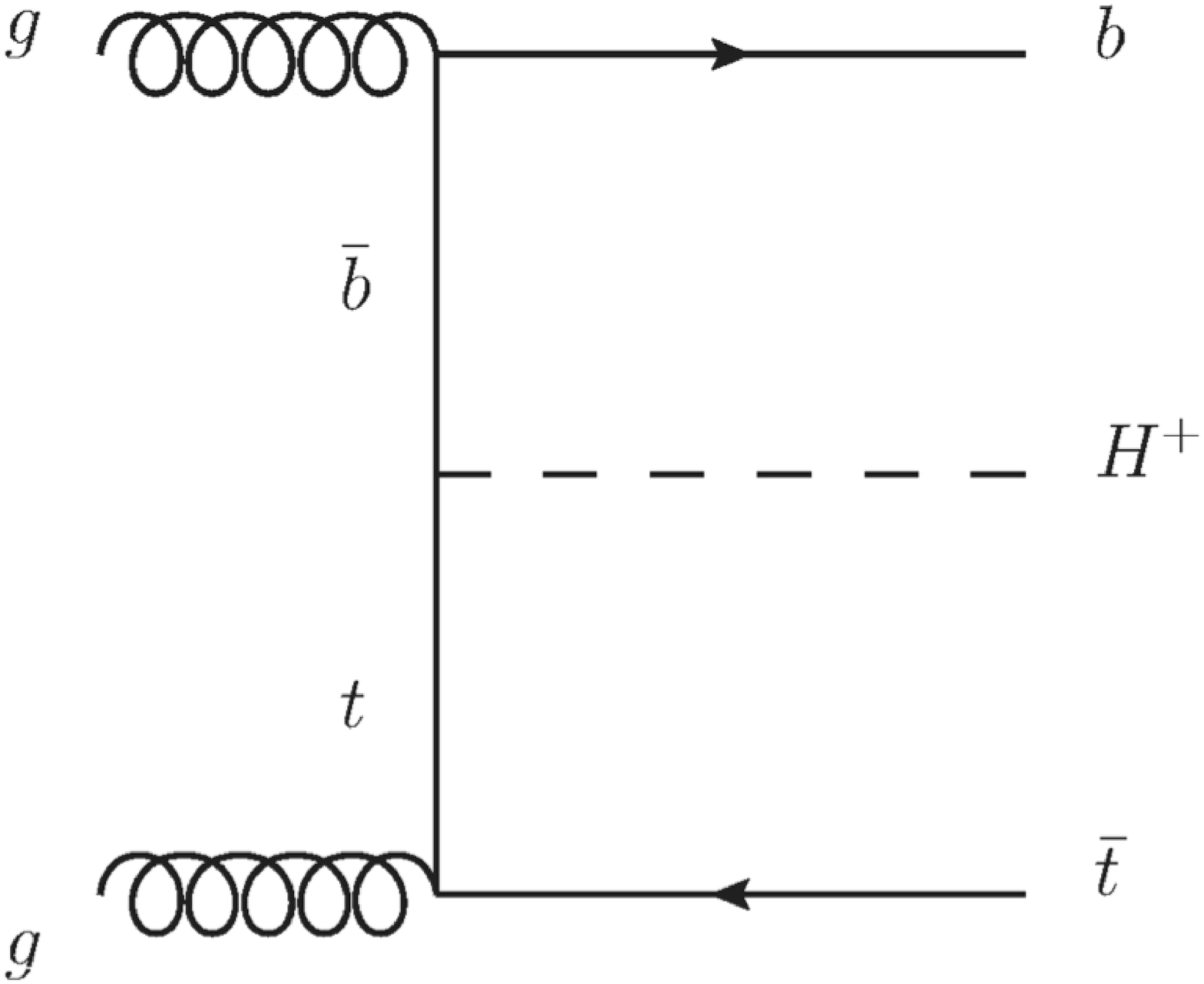}
}

\end{center}
\vspace*{-3mm}
\caption{Leading-order Feynman diagrams for the dominant production modes
of charged Higgs bosons at masses (a) below and (b, c) above
the top-quark mass.
\label{fig01}}
\end{figure}

Charged Higgs bosons ($H^{+}$, $H^{-}$) are predicted by several 
non-minimal Higgs scenarios, such as two-Higgs-doublet Models 
(2HDM)~\cite{Lee:1973iz} or models containing Higgs 
triplets~\cite{Cheng:1980,Schechter:1980,Lazarides:1981,Mohapatra:1983,Magg:1980}. 
As the Standard Model (SM) does not contain any elementary 
charged scalar particle, the observation of a charged Higgs 
boson\footnote{In the following, charged Higgs bosons are 
denoted by $H^+$, and the charge-conjugate is implied.} 
would clearly indicate new phenomena beyond the SM. 
For instance, supersymmetric models
predict the existence of charged Higgs bosons. In a type-II 2HDM, 
such as the Higgs sector of the Minimal Supersymmetric Standard Model (MSSM)~\cite{Fayet:1976et,Fayet:1977yc,Farrar:1978xj,Fayet:1979sa,Dimopoulos:1981zb}, 
the main $H^+$ production mode at the Large Hadron Collider (LHC) 
would be through top-quark decays ${t \rightarrow bH^+}$, for charged 
Higgs boson masses ($m_{H^+}$) smaller than the top-quark mass ($m_{\text{top}}$). 
At the LHC, top quarks are produced predominantly 
through $t\bar{t}$ production. 
In this paper, the contribution to ${t \rightarrow bH^{+}}$ which may arise from
single top-quark production is neglected, since the signal production cross section through this 
channel is very small with respect to $\ttbar$ production.
A diagram illustrating the leading-order production 
mechanism is shown on the left-hand side of figure~\ref{fig01}.
For charged Higgs boson masses larger than
$m_{\text{top}}$, the main $H^+$ source at the LHC
is through associated production with a top quark.
An additional $b$-quark can also appear in the final state.
The leading-order production mechanisms in two different approximations
are illustrated in the centre and right-hand side diagrams of figure~\ref{fig01}:
in the four-flavour scheme (4FS) $b$-quarks are dynamically produced,
whereas in the five-flavour scheme (5FS) the $b$-quark is also
considered as an active flavour inside the proton.
Their cross sections are matched according to ref.~\cite{Harlander:2011aa}, and an evaluation of 
the two schemes can be found in ref.~\cite{Aad:2014dvb}.
 
In the MSSM, the Higgs sector can be completely determined at tree level by one of the Higgs boson masses, here
taken to be $m_{H^+}$, and $\tan{\beta}$, the ratio of the vacuum expectation values of the two Higgs
doublets. For $m_{H^{+}}<m_{\text{top}}$, the decay via
${H^{+} \rightarrow \taup \nu}$ is dominant for $\tan\beta>2$ and remains 
sizeable for $1 < \tan\beta < 2$.
For higher $m_{H^+}$, the decay via ${H^+\rightarrow\taup\nu}$
is still significant, especially for large values of $\tan\beta$~\cite{Dittmaier:2011ti}.
The combined LEP 
lower limit for the charged Higgs boson mass is about 90\,GeV~\cite{:2001xy}. 
The Tevatron experiments placed upper limits on 
${{\cal B}(t \rightarrow bH^{+})}$ in the 15--20\% range for 
${m_{\text{H}^+}<m_{\text{top}}}$~\cite{Aaltonen:2009ke,:2009zh}.
In a previous search based on data taken at $\sqrt{s}=7$\,TeV with
the ATLAS and CMS detectors, the limits on ${{\cal B}(t \rightarrow bH^{+})}$
were lowered to the range 0.8--4\%~\cite{Aad:2012rjx,Chatrchyan:2012vca}.
For all of these results, ${{\cal B}(H^+ \rightarrow \taup\nu) = 100\%}$ was assumed.

This paper describes a search for charged Higgs bosons with masses 
in the ranges 80--160\,GeV\ and 180--1000\,GeV. 
The region 160\,GeV\ $ < m_{H^+} < 180 $\,GeV is not considered in this paper, since there is currently 
no reliable theoretical treatment for the interference between the different $\Hp$ production modes in this transition region~\cite{Berger:2003sm}.  
The final state studied is characterised by the presence
of a hadronic $\tau$ decay ($\tauhad$), missing transverse momentum ($\met$), $b$-quark-initiated jets, a
hadronically decaying $W$ boson, and the absence of any isolated electrons or muons with high transverse momenta.
In addition to the large branching ratio for a $\tau$ to decay hadronically,
this final state contains only neutrinos associated with the $H^+$ production and decay, resulting in good discriminating
power between SM and signal processes.
Charged Higgs bosons are 
searched for in a model-independent way, hence results  
are given in terms of ${{\cal B}(t\rightarrow bH^+)}\times {\cal B}(H^+ \rightarrow\taup\nu)$ (low-mass search, $m_{H^+}<m_{\text{top}}$)
and ${\xsection \times \brhtn}$ (high-mass search, $m_{H^+}>m_{\text{top}}$). 
These limits are then also interpreted in different MSSM scenarios.
The results are based on \LUMI 
of data from $pp$ collisions at ${\sqrt{s}=8~\mbox{TeV}}$, collected 
in 2012 with the ATLAS detector at the LHC.
The final state analysed for the low-mass search is 
${t\bar{t} \to b\bar{b}W^-H^+ \to b\bar{b}(q\bar{q}')(\tauhad^+ \nu)}$. 
The final state is similar or identical for
the high-mass search, depending on whether the additional $b$-quark-initiated jet is seen in the detector, $gb \to \bar{t}H^+ \to (W^-\bar{b})H^+ \to (q\bar{q}'\bar{b})(\tauhad^{+}\nu)$ in the 5FS 
case and 
$gg \to \bar{t}bH^+ \to (W^-\bar{b})bH^+ \to (q\bar{q}'\bar{b})b(\tauhad^{+}\nu) $ in the 4FS case.\\ 

This paper is organised as follows. In section~2, the data and simulated samples used in this analysis are 
described. In section~3, the reconstruction of physics objects in ATLAS 
is discussed. The event selection and background modelling are presented in section~4. 
Systematic uncertainties are discussed in section~5, 
and the limit-setting procedure is described in section~6.
Exclusion limits in
terms of ${{\cal B}(t\rightarrow bH^+)\times{\cal B}(H^+\rightarrow\taup\nu)}$ (low-mass) and ${\xsection \times \brhtn}$ (high-mass)
as well as model-dependent exclusion contours
are presented in section~7.

%%%%%%%%%%%%%%%%%%%%%%%%%%%%%%%%%%%%%%%%%%%%%%%%%%%%%%%%%%%%%%%%%%%%%%%%%%%%%%%
% Samples
%%%%%%%%%%%%%%%%%%%%%%%%%%%%%%%%%%%%%%%%%%%%%%%%%%%%%%%%%%%%%%%%%%%%%%%%%%%%%%%

\section{Data and simulated events}

The ATLAS detector \cite{Aad:2008zzm} consists of an inner tracking 
detector with coverage in pseudorapidity\footnote{
ATLAS uses a right-handed coordinate system with
its origin at the nominal interaction point (IP) in the centre
of the detector and the $z$-axis along the beam pipe. The $x$-axis
points from the IP to the centre of the LHC ring, and the $y$-axis
points upwards. Cylindrical coordinates $(r,\phi)$ are used in the
transverse plane, $\phi$ being the azimuthal angle around the $z$-axis.
The pseudorapidity is defined in terms of the polar angle $\theta$ as
$\eta=-\ln\tan(\theta/2)$.}
up to $|\eta|=2.5$, surrounded by a thin 2\,T superconducting solenoid, 
a calorimeter system extending up to $|\eta|=4.9$ 
and a muon spectrometer 
extending up to $|\eta|=2.7$ that measures the deflection of muon 
tracks in the field of three superconducting toroid magnets. A 
three-level trigger system is used. The first-level trigger (L1) is 
implemented in hardware, using a subset of detector information 
to reduce the event rate to no more than 75\,kHz. This 
is followed by two software-based trigger levels (L2 and EF), which together 
further reduce the event rate to less than 1\,kHz.

Only data taken with all ATLAS subsystems operational 
are used. Stringent detector and data quality requirements are applied,
resulting in an integrated luminosity of \LUMI for the 2012 data-taking 
period. The integrated luminosity has an uncertainty of 2.8\%,
measured following the methodology described in ref.~\cite{Aad:2013ucp}. 
Events are required to have a primary vertex with at least five associated tracks, 
each with a transverse momentum $\pt$ greater than 400 MeV. 
The primary vertex is defined as the reconstructed vertex with the largest sum of
squared track transverse momenta. 

The background processes to this search include SM pair production of top quarks, 
as well as the production of single top-quark, $W$+jets, $Z/\gamma^*$+jets,
diboson and multi-jet events. These backgrounds are categorised 
based on the type of reconstructed objects identified as the visible decay products
\footnote{This refers to the non-neutrino decay products.} of the hadronically decaying 
$\tau$ candidate ($\tauhadvis$). The dominant backgrounds in this analysis, those containing a true $\tauhad$, 
where the $\tauhadvis$ is correctly identified, or a jet misidentified as a $\tauhadvis$ candidate, 
are estimated in a data-driven way (sections~\ref{sec:embedding_taujets} and~\ref{sec:qcd_taujets}), while simulation 
samples are used to estimate the minor background arising from events with a lepton misidentified as a $\tauhadvis$ (1--2\% of the total 
background). Simulation samples are also used to develop and validate the analysis. 

The modelling of SM $t\bar{t}$ and single top-quark events is performed 
with MC@NLO~\cite{Frixione:2002ik, Frixione:2005vw}, except for $t$-channel single 
top-quark production, for which AcerMC~\cite{acer} is used. The top-quark 
mass is set to 172.5\,GeV\ and the set of parton distribution functions 
used is CT10~\cite{Lai:2010vv}. For events generated with MC@NLO, 
the parton shower, hadronisation and underlying event are added using 
HERWIG~\cite{hw} and JIMMY~\cite{jm}.  PYTHIA6~\cite{Sjostrand:2006za} 
is used instead for events generated with AcerMC. Inclusive cross 
sections are taken from the approximate next-to-next-to-leading-order (NNLO) predictions 
for \ttbar\ production~\cite{hathor}, for single top-quark production in the 
$t$-channel and $s$-channel~\cite{Kidonakis:2011wy,Kidonakis:2010tc}, as well 
as for $Wt$ production~\cite{Kidonakis:2010ux}.
Overlaps between SM $Wt$ and $t\bar{t}$ final states are 
removed~\cite{Frixione:2005vw}.
Single vector boson ($W$ and $Z/\gamma^*$) production is 
simulated with up to five accompanying partons, using ALPGEN~\cite{Mangano:2002ea} interfaced to HERWIG 
and JIMMY, and using the CTEQ6L1~\cite{CTEQ6L1} parton distribution functions. 
The additional partons produced in the matrix-element part of the event 
generation can be light partons or heavy quarks. In the latter case, 
ALPGEN is also used to generate dedicated samples with matrix elements for the production of massive
$b\bar{b}$ or $c\bar{c}$ pairs.
Diboson events ($WW$, $WZ$ 
and $ZZ$) are generated using HERWIG. The cross sections are 
normalised to NNLO predictions for $W$-boson and $Z/\gamma^*$ 
production~\cite{Gavin:2012kw,Gavin:2010az} and to 
next-to-leading-order (NLO) predictions for diboson 
production~\cite{Campbell:2011bn}. The SM background samples are summarised in 
table~\ref{xs}. 

Signal samples are 
produced with PYTHIA 6 for 80\,GeV\ $ \leq m_{H^+} \leq 160$\,GeV in $m_{H^+}$ intervals of 10\,GeV separately for 
${t\bar{t} \rightarrow b\bar{b}H^+W^-}$ and 
${t\bar{t} \rightarrow b\bar{b}H^-W^+}$, where the
charged Higgs bosons decay via ${H^+ \rightarrow \taup\nu}$. The process 
${t\bar{t} \rightarrow b\bar{b}H^+H^-}$ gives a very small contribution to 
the signal region, which is negligible after the event selection described in 
section~\ref{sec:event_taujets}. 
The cross section for these processes depends 
only on the total $t\bar{t}$ production cross section
and the branching ratio ${\cal B}(t \to bH^+)$.
For 180\,GeV\ $ \leq m_{H^+} \leq 1000$\,GeV, the simulation of the signal for top-quark associated
$H^{+}$ production is performed with POWHEG~\cite{Frixione:2007vw} interfaced to PYTHIA 8~\cite{Sjostrand:2007gs}.
For 180\,GeV\ $ \leq m_{H^+} \leq 200$\,GeV, samples are produced in $m_{H^+}$ steps of 10\,GeV, then
in intervals of 25\,GeV up to $m_{H^+}=300$\,GeV and in intervals of 50\,GeV for $m_{H^+}\leq 600$\,GeV.
Additionally, signal mass points at $m_{H^+}=750$\,GeV and $m_{H^+}=1000$\,GeV are produced.
The production cross section for the high-mass charged Higgs boson is computed using
the 4FS and 5FS, including theoretical uncertainties, and combined 
according to ref.~\cite{Harlander:2011aa}. The samples are generated at NLO using the 5FS and the narrow-width
approximation for the $\Hp$.  Possible effects from the interference
between the production of a charged Higgs boson through $\ttbar$ and top-quark associated production 
are not taken into account. 

\begin{table}[h!]
  \begin{center}
    \begin{tabular}{l|llr}
Process   &  Generator & Cross section [pb]& \\ 
\hline \hline
SM $t\bar{t}$ (inclusive) & 
MC@NLO & 
$253\phantom{.0 \times 10^0}$ & \cite{hathor}  \\ 
\hline
Single top-quark $t$-channel ($\geq$ 1 lepton) & AcerMC &
$\phantom{0}28.4\phantom{.0 \times 10^0}$&  \cite{Kidonakis:2011wy} \\ 
Single top-quark $s$-channel ($\geq$ 1 lepton) & MC@NLO &
$\phantom{00}1.8\phantom{\times 10^0}$ & \cite{Kidonakis:2010tc} \\
Single top-quark $Wt$-channel (inclusive) & MC@NLO & 
$\phantom{0}22.4\phantom{.0\times 10^0}$ & \cite{Kidonakis:2010ux} \\ 
\hline
$W \rightarrow \ell\nu$ & ALPGEN & 
$\phantom{00}3.6 \times 10^4$ &\cite{Gavin:2012kw} \\ 
\hline
$Z/\gamma^* \rightarrow \ell\ell$ with $m(\ell\ell) > 10~\mbox{\GeV}$ & 
ALPGEN & 
$\phantom{00}1.7 \times 10^4$ & \cite{Gavin:2010az} \\
\hline
$WW$ ($\geq$ 1 electron/muon) & HERWIG & 
$\phantom{0}20.9\phantom{.0\times 10^0}$ & \cite{Campbell:2011bn} \\ 
$ZZ$ ($\geq$ 1 electron/muon) & HERWIG & 
$\phantom{00}1.5 \phantom{\times 10^0}$ & \cite{Campbell:2011bn} \\
$WZ$ ($\geq$ 1 electron/muon) & HERWIG & 
$\phantom{00}7.0 \phantom{\times 10^0}$ & \cite{Campbell:2011bn} \\
\hline
$H^+$ signal ($m_{H^+}=250$\,GeV)   &  POWHEG & $\phantom{00}0.5$ & \\
    \end{tabular}
\caption{Cross sections for the simulated processes and reference generators used to model them.
For the high-mass $H^+$ signal selection, the value shown is the cross section times ${{\cal B}(H^+ \rightarrow \taup \nu)}$ 
for the MSSM $m_h^{\text{max}}$ scenario \cite{Carena:1999xa, Carena:2013qia}, corresponding to $m_{H^+} =$ 250\,GeV\ 
and $\tan{\beta} = 50$. This cross section includes both $H^+$ and $H^-$ production. The low-mass signal, which is not 
included in the table, assumes one $H^{+}$ produced per $t \bar{t}$ decay, so it is a fraction of the $\ttbar$ cross section. 
The previously published upper limit on ${\cal B}(t \to bH^+)$ for $m_{H^+} =$ 130\,GeV\ is 0.9$\%$~\cite{Aad:2012rjx}.
\label{xs}}
  \end{center}
\end{table}

The event generators are tuned to describe the ATLAS data.
In samples where PYTHIA 6 is interfaced to AcerMC,
the AUET2B~\cite{pythiatune} tune is used.
The Perugia 2011 C tune~\cite{isrfsr} is used when PYTHIA 6 is interfaced to
POWHEG. For the samples generated with HERWIG,
the AUET2~\cite{herwigtune} tune is used.
In all samples with $\tau$ leptons, except for those simulated with PYTHIA 8\footnote{So-called ``sophisticated tau-decays'' 
have been available in PYTHIA since version 8.150 such that the usage of TAUOLA is not necessary.},
TAUOLA~\cite{Was:2004dg} is used for the $\tau$ decays.
PHOTOS~\cite{Barberio:1990ms} is used for photon radiation from charged leptons in all samples
where applicable.

To take into account the presence of multiple proton--proton interactions 
occurring in the same and neighbouring 
bunch crossings (referred to as pile-up), 
simulated minimum-bias events are added to the hard process 
in each generated event. Prior to the analysis, simulated events are 
reweighted in order to match the distribution of the average number of 
pile-up interactions in the data. All generated events are propagated 
through a detailed GEANT4 simulation~\cite{Agostinelli:2002hh,simu} of 
the ATLAS detector and are reconstructed with the same algorithms as the data.

%%%%%%%%%%%%%%%%%%%%%%%%%%%%%%%%%%%%%%%%%%%%%%%%%%%%%%%%%%%%%%%%%%%%%%%%%%%%%%%
% Atlas
%%%%%%%%%%%%%%%%%%%%%%%%%%%%%%%%%%%%%%%%%%%%%%%%%%%%%%%%%%%%%%%%%%%%%%%%%%%%%%%

\section{Physics object selection}
\label{sec:object}

Jets are reconstructed from energy deposits in the calorimeters, using the anti-$k_t$ 
algorithm~\cite{Cacciari:2008gp, Cacciari:2005hq} 
with a radius parameter of \mbox{$R = 0.4$}. Jets are required
to have $\pt >25$\,GeV\ and $|\eta| < 2.5$. To reduce the contribution of jets initiated by pile-up, 
jets with ${\pt < 50}$\,GeV and $|\eta| <$ 2.4 must pass the requirement that at least half of the 
$\pt$ of the tracks associated with the jet is contributed by tracks matched to the primary vertex~\cite{Abazov:2006yb}. 
An algorithm identifies jets containing $b$-quarks by 
combining impact parameter information with the explicit 
determination of a secondary vertex~\cite{ATLAS:2012ima}, and these are referred 
to as $b$-tagged jets. A working
point corresponding to a $70\%$ efficiency for identifying $b$-quark-initiated jets is used. 

Candidates for identification as $\tauhadvis$ arise from jets reconstructed 
from energy deposits in calorimeters, again using the anti-$k_t$
algorithm with a radius parameter of \mbox{$R = 0.4$}, which have
$\pt>10$\,GeV\ and one or three charged-particle tracks within 
a cone of size of $\Delta R<0.2$, where $\Delta R =\sqrt{(\Delta \eta)^2+ (\Delta\phi)^2}$ 
around the $\tauhadvis$ axis~\cite{taus}. 
These candidates are further required to 
have a visible transverse momentum ($\pt^\tau$) of at
least 20\,GeV\ and to be within $|\eta| < 2.3$. The output of boosted decision tree
algorithms~\cite{breiman1984classification, Freund1997119} is used to distinguish  
$\tauhadvis$ from jets not initiated by $\tau$ leptons, separately
for $\tauhad$ decays with one or three charged-particle tracks. 
In this analysis, a point with 40$\%$ (35$\%$)
efficiency for identification of 1(3)-prong $\tauhadvis$ is used, and this requirement is
referred to as the $\tauhadvis$ identification.  
Dedicated algorithms are used to reject electrons and muons that are
incorrectly identified as $\tauhadvis$~\cite{taus}. After 
these algorithms are applied, the backgrounds arising from muons and 
electrons misidentified as $\tauhadvis$ are very small,
although there is still a sizeable background from jets misidentified as $\tauhadvis$.

The $\met$ is defined as the magnitude of the negative vectorial sum
of transverse momenta of muons and energy deposits in the calorimeter. It is computed using fully calibrated
and reconstructed physics objects~\cite{MET2013}.

The final states considered 
in this search contain no charged leptons, hence events containing isolated electron 
or muon candidates with high transverse momenta are rejected. Electron candidates are reconstructed from
energy deposits in the calorimeter that are matched to tracks in the inner detector,
taking losses due to bremsstrahlung into account.
They are required to have a transverse energy ($\ET$) greater than 25\,GeV\ and to be within $|\eta|<2.47$ (the transition
region between the barrel and end-cap calorimeters, $1.37<|\eta|<1.52$,
is excluded)~\cite{ATLAS-CONF-2014-032,egamma_new}. Muon candidates must pass tracking requirements in both the inner detector
and the muon spectrometer, have $\pt > 25$\,GeV and $|\eta| < 2.5$~\cite{sys_muon_rec_eff}. Additionally, electron candidates 
are required to pass pile-up-corrected 90$\%$ efficient calorimeter- and track-based isolation, with $\Delta$R cone 
sizes of 0.2 and 0.3, respectively, while muon candidates are required to pass a relative track-based isolation of $<$ 0.05 with 
a $\Delta$R cone $<$ 0.4~\cite{ATLAS:2014aga}.

%%%%%%%%%%%%%%%%%%%%%%%%%%%%%%%%%%%%%%%%%%%%%%%%%%%%%%%%%%%%%%%%%%%%%%%%%%%%%%%
% tau+jets
%%%%%%%%%%%%%%%%%%%%%%%%%%%%%%%%%%%%%%%%%%%%%%%%%%%%%%%%%%%%%%%%%%%%%%%%%%%%%%%

\section{Event selection and background modelling}\label{sec:taujets}

\subsection{Event selection}
\label{sec:event_taujets}

The analysis uses events passing a $\tauhadvis$+$\met$ trigger. The $\tauhadvis$ trigger is
 defined by calorimeter energy in a narrow core region and an isolation region at L1, 
a basic combination of tracking and calorimeter information at L2 and 
more sophisticated algorithms imported from the offline reconstruction at the EF. 
The $\met$ trigger uses calorimeter information at all levels with a more refined algorithm at the EF.
The EF threshold on the transverse momentum of the $\tauhadvis$ trigger object is 27\,GeV or 29\,GeV, 
and for the $\met$ trigger the EF threshold is 40\,GeV or 50\,GeV. The multiple trigger thresholds are 
the result of slight changes of the trigger definition during the 2012 data-taking period, for which 50$\%$ 
of events had EF thresholds at 27\,GeV and 50\,GeV, 43$\%$ at 29\,GeV and 50\,GeV, and 7$\%$ at 29\,GeV and 40\,GeV, 
for the $\tauhadvis$ and $\met$ triggers, respectively. 

Further event filtering is
performed by discarding events in which any jet with $\pt>25$\,GeV\ fails
the quality cuts discussed in ref.~\cite{atlas_event_cleaning}.
This ensures that no jet is consistent with having originated
from instrumental effects
or non-collision backgrounds. The following requirements are then applied:
\begin{itemize}
  \item at least four (three) selected jets for the low-mass (high-mass) signal selection;
  \item at least one of these selected jets being $b$-tagged at the 70\%-efficient working point;
  \item exactly one selected $\tauhadvis$ with $\pt^\tau > 40$\,GeV\ matched to a $\tauhadvis$ trigger object (trigger-matched); 
  \item no selected electron or muon in the event;
  \item $\met > 65$ (80)\,GeV\ for the low-mass (high-mass) signal selection;
    \item $\met / \sqrt{\sum{\pT^{\text{PV trk}}}} >$ 6.5 (6.0)\,GeV$^{1/2}$ for the low-mass (high-mass) signal selection,
    where $\sum{\pT^{\text{PV trk}}}$ is the sum of transverse momenta of all tracks originating from the primary vertex.
    This is to reject events in which a large reconstructed \met~is due to the limited resolution of the energy
    measurement. 
\end{itemize}

For the selected events, the transverse mass ($\mt$) of the $\tauhadvis$ and $\met$ is defined as:
\begin{equation}
  \label{eq:mt}
\mt=\sqrt{ 2 \pt^\tau  \ET^{\mathrm{miss}} (1-\cos \Delta\phi_{\tau,\text{miss}}) },
\end{equation}
where $\Delta\phi_{\tau,\text{miss}}$ is the azimuthal angle between the 
$\tauhadvis$ and the direction of the missing transverse momentum. This discriminating 
variable takes values lower than the $W$ boson mass 
for $W\rightarrow\tau\nu$ background events
and less than the $H^+$ mass for signal events, in the absence of detector 
resolution effects. 

A minimal requirement is placed on $\mt$ at 20 (40)\,GeV\ in the low-mass (high-mass) $H^+$ search.
This requirement is motivated in section~\ref{sec:embedding_taujets}.  After the full event selection, 
the signal has an acceptance of 0.30--0.60$\%$ for the low-mass range, and 1.7--5.8$\%$ for the high-mass range, 
where in both cases the acceptance increases with increasing $m_{H^+}$.
The acceptances are evaluated with respect to signal samples where both the $\tau$ lepton and the associated top quark decay inclusively.

\subsection{Data-driven estimation of the backgrounds with a true $\tauhad$}
\label{sec:embedding_taujets}

An embedding method~\cite{Aad:2011rv} 
is used to estimate the backgrounds that contain a real $\tauhad$ from a vector boson decay.
The method is based on a control data sample 
of $\mu$+jets events 
satisfying criteria similar to those of
the signal selection
except for the $\tauhadvis$ requirements
and replacing the detector signature 
of the muon by a simulated hadronic $\tau$ decay. The method is applied to a 
control region of $\mu$+jets events, rather than $e$+jets, due to the clean signature and the
relative ease with which the measured muon can be removed. These new hybrid events 
are then used for the background prediction.
An advantage of this approach, compared to simulation, is that with the exception 
of the $\tauhad$, the estimate is extracted from data;
this includes the 
contributions from the underlying event and pile-up, jets, 
and all sources of $\met$ except for the neutrino from the $\tauhad$ decay.
Furthermore, since the normalisation of the background estimate is evaluated
from the data, assuming lepton universality of the $W$ boson decay, the method does not rely on theoretical
cross sections and their uncertainties. 
This embedding method has been used in previous charged Higgs
boson searches~\cite{Aad:2012tj} as well as in SM ${H\rightarrow\tau\tau}$~\cite{Aad:2012mea,Aad:2015vsa} analyses.

To select the $\mu+$jets
sample from the data, the following requirements are made:
\begin{itemize}
\item a single-muon trigger with a $\pt$ threshold of $24$\,GeV\ or $36$\,GeV (single-muon triggers with two different $\pt$ thresholds are used, since the lower-threshold trigger
also requires the muon to be isolated);
\item exactly one isolated muon with $\pt >25$\,GeV and no isolated electron with $\phantom{000000000}$ $\ET>25$~GeV;
\item at least four (three) jets with $\pt >25$\,GeV for the low-mass
(high-mass) charged Higgs boson search, at least one of which is $b$-tagged;
\item $\ET^{\mathrm{miss}}>25$ (40)\,GeV\ for the low-mass (high-mass) charged Higgs
boson search.
\end{itemize}

This selection is looser than the selection defined in 
section~\ref{sec:event_taujets} in order not to bias the sample. 
However, the $\met$ cut in the $\mu+$jets sample selection removes 
events with very low $\mt$. Thus, a cut on $\mt>20$ (40)\,GeV\ is introduced in the search
for low-mass (high-mass) charged Higgs bosons to remove this bias. With this selection, there 
is a possible small contamination from signal events with a leptonically decaying $\tau$ lepton. 
This small contamination, which is estimated using simulation, has a much softer $\mt$ distribution than the signal with $\tauhad$, and 
is observed to have a negligible impact on the evaluation of signal strength or exclusion limits. Contamination from
leptonically decaying $\tau$ leptons from $W$ decays is accounted for in the overall normalisation ($c_{\tau\to\mu}$ in eq.~(\ref{eqn:tau_embed_norm})).

To replace a muon in the selected data, the track that is associated
with the muon is removed. The energy deposited in the calorimeters is removed
by simulating a ${W\rightarrow\mu\nu}$ event with the same kinematics as in
the selected data event and identifying the corresponding cells.
Thus, the removal of energy deposits not associated with the selected muon is minimised.
The momentum of the muon in selected events is extracted and rescaled
to account for the higher $\tau$ lepton mass, 
\begin{equation}
  \vec{p}_{\tau} =\frac{\sqrt{E_{\mu}^{2}-m_{\tau}^{2}}}{\sqrt{\vec{p}_{\mu}\cdot\vec{p}_{\mu}}}\vec{p}_{\mu},
\end{equation}
where $\vec{p}_{\tau}$ is the rescaled momentum, $E_{\mu}$ is the reconstructed energy of the muon,
$m_{\tau}$ is the $\tau$ mass, and $\vec{p}_{\mu}$ is the reconstructed muon momentum.
The $\tau$ lepton with rescaled momentum is further processed by TAUOLA to produce the hadronic
$\tau$ decay and account for the $\tau$ polarisation as well as for final-state radiation. The $\tau$ lepton decay
products are propagated through the full detector simulation and reconstruction.
Events referred to as containing a true $\tauhad$ are those with a genuine $\tauhad$ as expected from the embedding method.

The shape of the $\mt$ distributions for backgrounds with a true $\tauhad$ 
is taken from the distribution obtained with the embedded events, 
after applying the corresponding signal selection. The 
normalisation is then derived from the number of embedded events: 
\begin{equation}
  \label{eqn:tau_embed_norm}
N_{\tau} = N_{\mathrm{embedded}} \cdot \left( 1 - c_{\tau\to\mu} \right) 
\frac{\epsilon^{\tau+\met\mathrm{-trigger}}}{\epsilon^{\mu\mathrm{-ID, trigger}}} \times {\cal B}(\tau \rightarrow \text{hadrons}+\nu),
\end{equation}
where $N_\tau$ is the estimated number of events with a true 
$\tauhad$, $N_{\mathrm{embedded}}$ is the number 
of embedded events in the signal region, $c_{\tau\to\mu}$ is the 
fraction of events in which the selected muon is a decay product 
of a $\tau$ lepton (taken from simulation, about $4\%$), 
$\epsilon^{\tau+\met\mathrm{-trigger}}$ is the 
$\tauhadvis{+}\met$ trigger efficiency (as a function of 
$\pt^\tau$ and $\met$, derived from data, see section~\ref{sec:trigger}), 
$\epsilon^{\mu\mathrm{-ID, trigger}}$ is the muon trigger and identification 
efficiency (as a function of $p_T$ and $\eta$, derived from data) and 
${{\cal B}(\tau \rightarrow \text{hadrons}+\nu)}$ is the branching ratio of 
the $\tau$ lepton decays to hadrons.

The $m_{\text{T}}$ distributions for selected events with a true $\tauhad$, 
as obtained with the embedding method, are shown in figure~\ref{fig:taujets_emb} 
and compared to simulation. Embedded data and simulation agree well and are within
uncertainties. The combined systematic and statistical uncertainties on the embedded 
prediction and simulation are compared directly in figure~\ref{fig:taujets_emb}, 
where the reduction provided by the use of the embedding method is shown.
 
\begin{figure}[h!]
\begin{center}
\subfigure[\label{fig:taujets_emb} Low-mass $H^+$ selection]{
\includegraphics[width=0.45\textwidth]{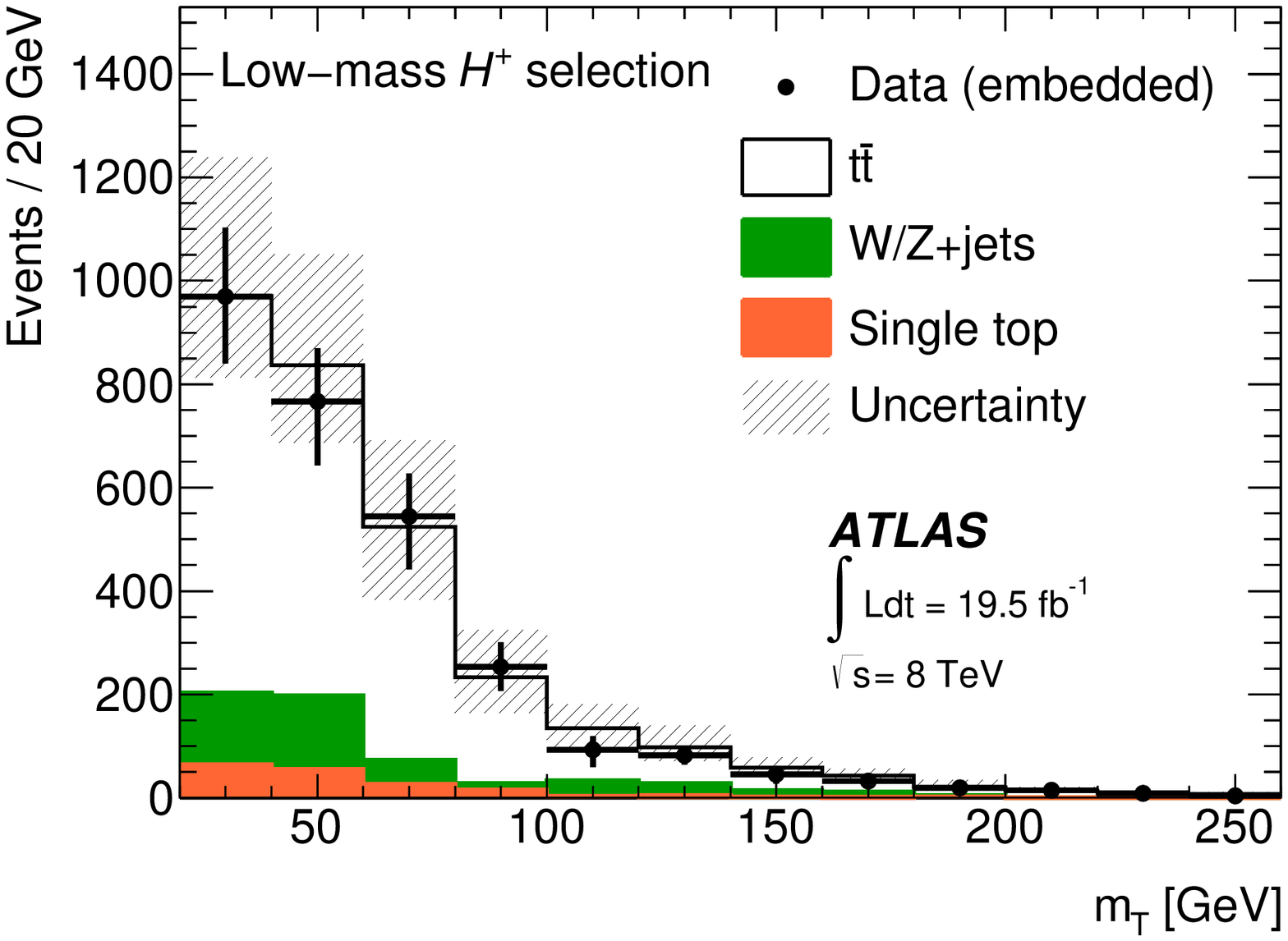}
}
\subfigure[\label{fig:taujets_emb} High-mass $H^+$ selection]{
\includegraphics[width=0.45\textwidth]{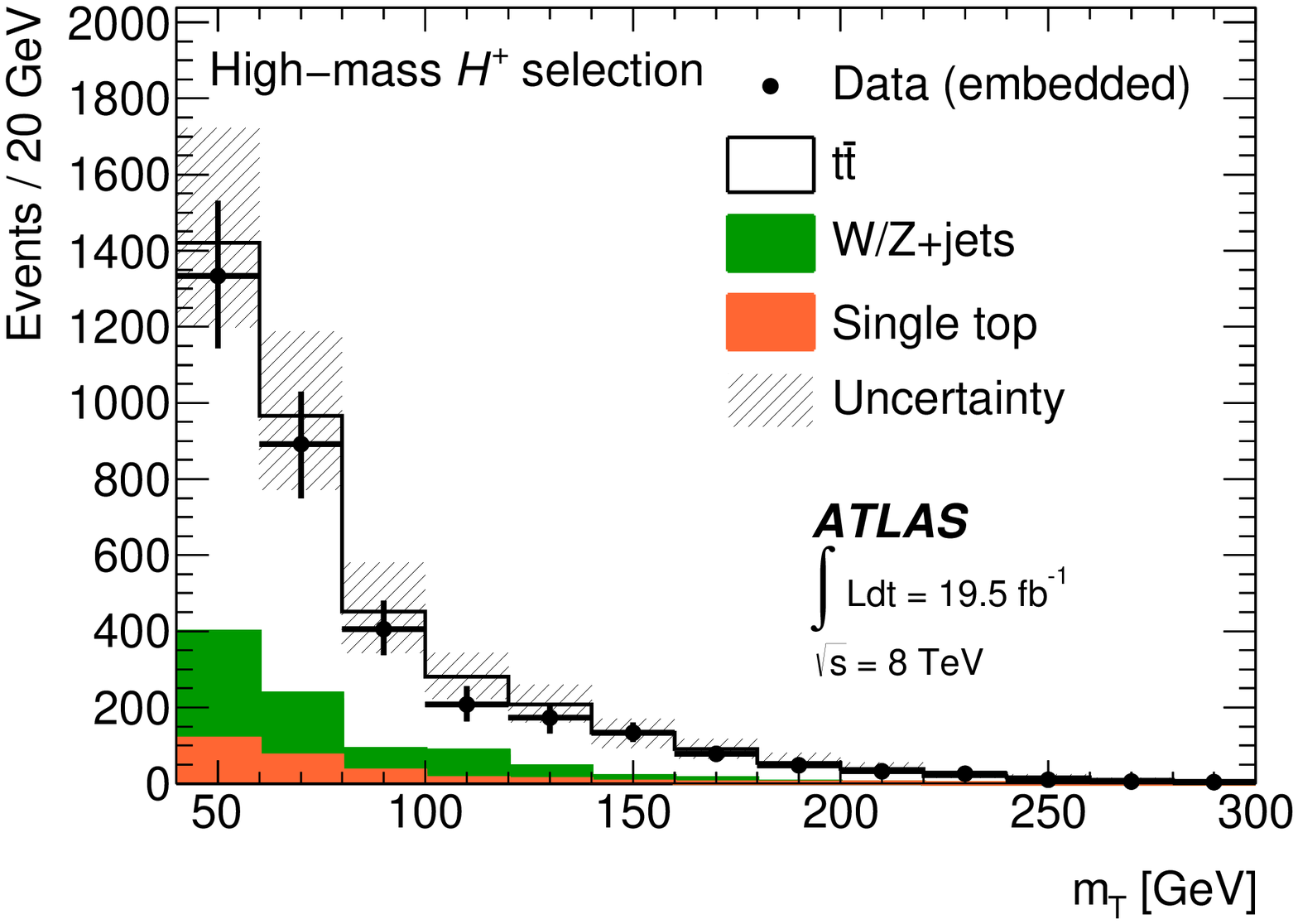}
}
\end{center}
\caption{\label{fig:taujets_emb}
Comparison of the $m_{\text{T}}$ distributions for events with a true $\tauhad$
for the (a) low-mass and (b) high-mass charged Higgs boson search, 
as predicted by the embedding method and simulation. 
Combined statistical and systematic uncertainties (as described in 
section~\ref{sec:systematics}) for the embedded sample are shown as error bars, and
all systematic uncertainties applicable to simulation are shown as hatched bands.
}
\end{figure}

\subsection{Data-driven estimation of the multi-jet backgrounds}
\label{sec:qcd_taujets}

For the data-driven estimation of the backgrounds with a jet misidentified as a $\tauhadvis$ (multi-jet background), two data
samples are defined, differing only in $\tauhadvis$ identification criteria. The \emph{tight} sample contains a larger fraction of 
events with a real $\tauhadvis$, which are required to pass the \emph{tight} $\tauhadvis$ identification selection described in
the object selection, in addition to the trigger matching required in the event selection of section~\ref{sec:event_taujets}. 
The \emph{loose} sample, which contains a larger fraction of events with a 
misidentified $\tauhadvis$, is obtained by removing the $\tauhadvis$ identification requirement that was applied
in the \emph{tight} sample. 
By construction, the \emph{tight} data sample is a subset of the \emph{loose} data sample.

The \emph{loose} 
sample consists of $N_{\text{r}}$ and $N_{\text{m}}$ events 
with, respectively, a real or misidentified $\tauhadvis$. It is also 
composed of $N_{\text{L}}$ events with a $\tauhadvis$ passing a 
\emph{loose but not tight} selection, and $N_{\text{T}}$ events in 
which the $\tauhadvis$ fulfils the \emph{tight} selection. 
Using the efficiencies $p_{\text{r}}$ and $p_{\text{m}}$, respectively, for 
a real or misidentified \emph{loose} $\tauhadvis$ satisfying the \emph{tight}
criteria, the following relation can be established: 

\begin{equation}
\left(
\begin{array}{c}
N_{\text{T}} \\
N_{\text{L}}
\end{array}
\right)
=
\left(
\begin{array}{cc}
p_{\text{r}} & 
p_{\text{m}} \\
(1-p_{\text{r}}) & 
(1-p_{\text{m}}) \\
\end{array}
\right)
\times
\left(
\begin{array}{c}
N_{\text{r}} \\
N_{\text{m}}
\end{array}
\right).
\end{equation}

In turn, inverting the $2 \times 2$ matrix above, the number of 
events in which the misidentified $\tauhadvis$ passes the \emph{tight} selection can be 
written as:
\begin{equation} 
\label{tight1}
N_{\text{m}}^{\text{T}} = p_{\text{m}}N_{\text{m}} =
\dfrac{p_{\text{m}}p_{\text{r}}}{p_{\text{r}}-p_{\text{m}}}N_{\text{L}} + 
\dfrac{p_{\text{m}}(p_{\text{r}}-1)}{p_{\text{r}}-p_{\text{m}}}N_{\text{T}}.
\end{equation} 
The final values of $p_{\text{r}}$ and $p_{\text{m}}$ are parameterised in terms of the number
of charged-particle tracks in the core cone ($\Delta R\leq 0.2$) and the number of charged-particle tracks in the hollow isolation cone ($0.2<\Delta R <0.4$) 
around the $\tauhadvis$ axis~\cite{taus}, 
as well as the $\pt$ and $|\eta|$ of the $\tauhadvis$. Correlations 
between the variables used for parameterisation are found to have a negligible effect on the results of the method.

The probability $p_{\text{r}}$ is determined using true $\tauhadvis$ in simulated $\ttbar$ events
in the signal region. The probability $p_{\text{m}}$ is measured in a $W$+jets control region in data. Events in this control region are triggered by a
combined trigger requiring an electron with $\et>18$\,\GeV\ or a muon with $\pt >15$\,\GeV\ in addition to a $\tauhadvis$. 
In both cases, the $\tauhadvis$ trigger object has a $\pt$ threshold of $20$\,GeV.  The control
region must have exactly one trigger-matched reconstructed electron or muon, in addition to a trigger-matched, reconstructed, \emph{loose} $\tauhadvis$. 
The control region is also required to have zero $b$-tagged jets and $\mt(e/\mu,\met) > 50$\,GeV (using eq.(~\ref{eq:mt}), with the $\tauhadvis$ replaced by the electron or muon). 
The contamination from correctly reconstructed $\tauhadvis$ ($7\%$) and electrons or muons mis-reconstructed as $\tauhadvis$ ($5\%$) is subtracted using
simulation. Signal processes contribute negligibly to this region ($< 0.1 \%$). 
 
Having computed the identification and misidentification efficiencies 
$p_{\text{r}}$ and $p_{\text{m}}$, every event in the \emph{loose} sample is 
given a weight $w$ as follows, in order to estimate the background with 
a misidentified $\tauhadvis$ in the \emph{tight} sample:
\begin{itemize}
\item for an event with a \emph{loose but not tight} $\tauhadvis$, 
$w_{\text{L}} = \displaystyle{{p_{\text{m}}p_{\text{r}} \over p_{\text{r}} - p_{\text{m}}}}$;
\item for an event with a \emph{tight} $\tauhadvis$, 
$w_{\text{T}} = \displaystyle{{p_{\text{m}}(p_{\text{r}}-1) \over p_{\text{r}} - p_{\text{m}}}}$.
\end{itemize}
 
Events with jets misidentified as $\tauhadvis$ are a major background 
in the high-$\mt$ region ($>300$\,\GeV\ low-mass and $>400$\,\GeV\ high-mass), but this region has less than one expected event per 20\,\GeV\ bin.
This limitation is circumvented by fitting the $\mt$ distribution using a power-log function 
in the mass range 200--800\,GeV.  The power-log function is defined by the following formula:

\begin{equation}
  f(x) = x^{a + b \ln(x)}, 
\end{equation}
where $a$ and $b$ are fitted constants.
The resulting $\mt$ distribution after considering each systematic uncertainty is fitted separately.
An additional systematic uncertainty is added for the choice of fit function, by symmetrising the 
difference between the baseline fit and an alternative fit using an exponential function.  The
exponential is chosen to probe the effect on the expected yield in the poor statistics tail region, 
since it also describes the multi-jet background well in the region with many events.  Figure~\ref{fig:qcd_extrapolation} shows 
the fits obtained in the nominal case, for the systematic uncertainty due to the chosen fit function, and for all other 
systematic uncertainties related to this background estimation (see section~\ref{sec:backgroundsystematics}).

\begin{figure}[h!]
\centering
\subfigure[\label{fig:qcd_extrapolation} Low-mass $H^+$ selection]{
\includegraphics[width=0.45\textwidth]{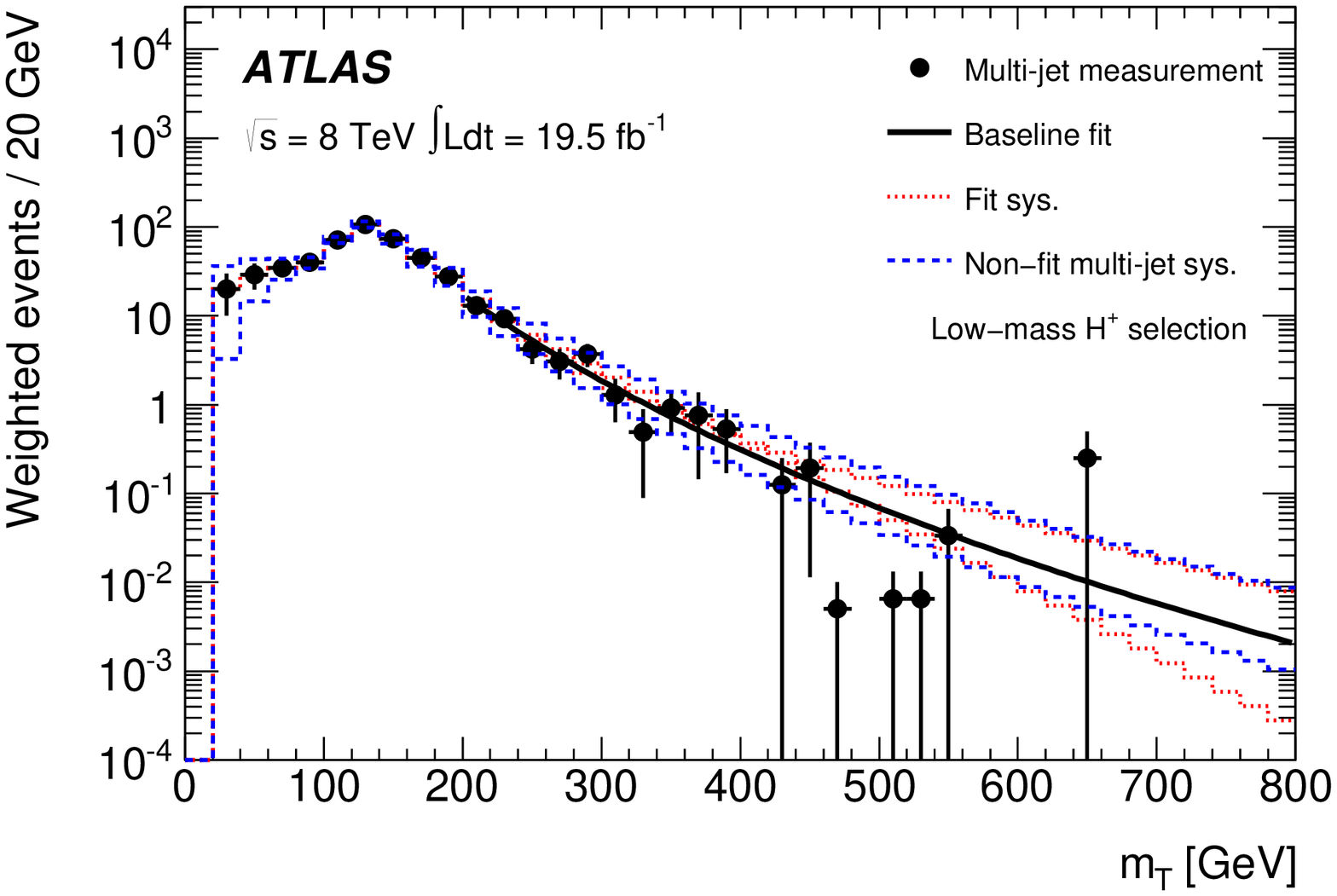}
}
\subfigure[\label{fig:qcd_extrapolation} High-mass $H^+$ selection]{
\includegraphics[width=0.45\textwidth]{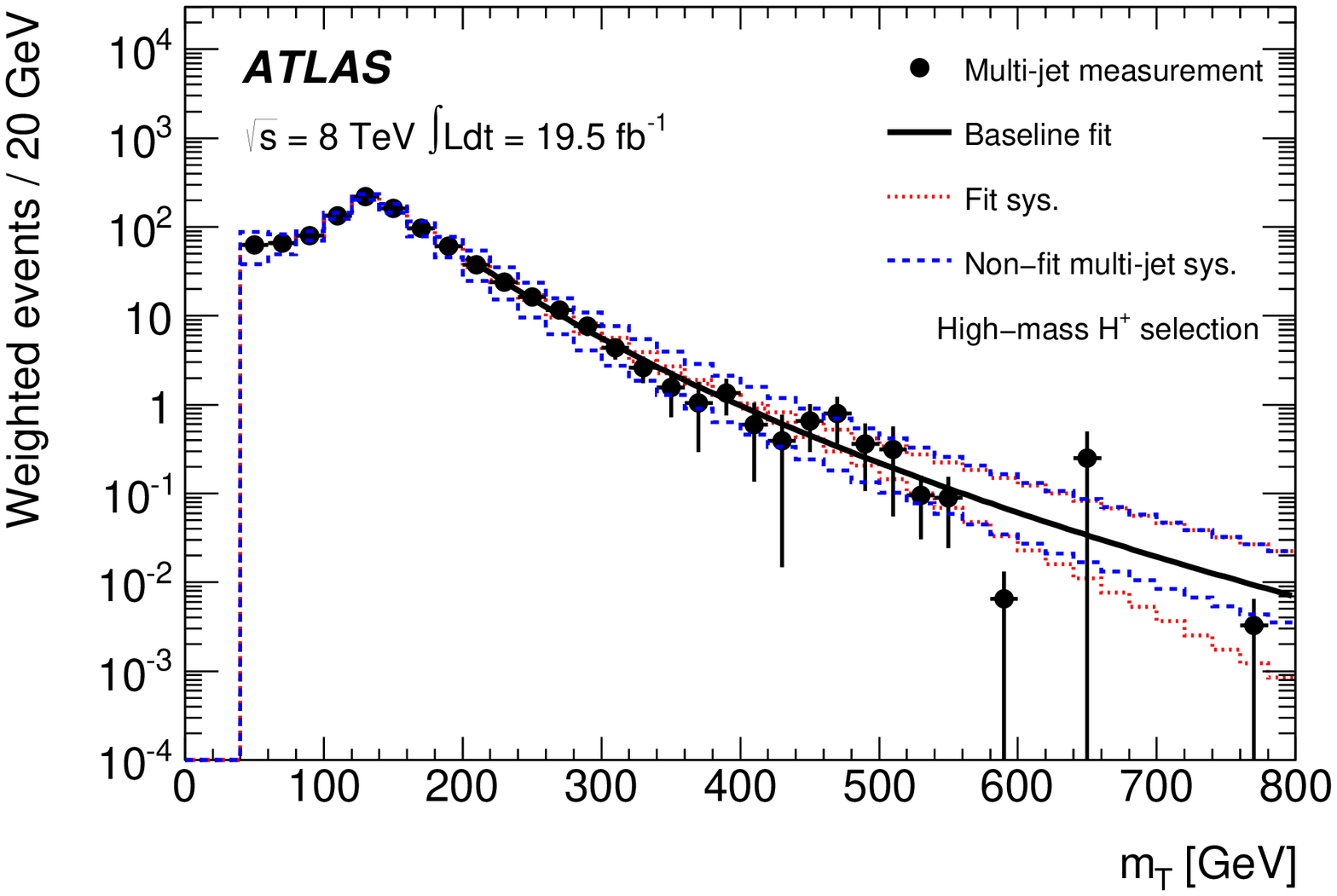}
}
\caption{\label{fig:qcd_extrapolation} 
The multi-jet background predictions from data-driven methods for the (a) low-mass and (b) high-mass $H^+$ event selections, 
with the results of fits using the
power-log function, are shown in the solid line. The dotted lines
show the systematic uncertainty from the choice of the fit function.  The dashed lines show the total combined fits from 
the sources of systematic uncertainty listed in table~\ref{tab:systmethod}.}
\end{figure}

\subsection{Backgrounds with electrons or muons misidentified as $\tauhadvis$}
\label{sec:taufakes_taujets}

Backgrounds that arise from events where an electron or muon is misidentified as $\tauhadvis$ are heavily 
suppressed by dedicated veto algorithms, so that these events only contribute
at the level of 1--2$\%$ to the total background.  These backgrounds are estimated from simulated events, and they 
include contributions from $\ttbar$, single top-quark, diboson, $W$+jets and $Z$+jets processes.  Leptons from in-flight
 decays in multi-jet events are accounted for in the multi-jet background estimate.

\subsection{$\tauhadvis{+}\met$ triggers}
\label{sec:trigger}
The analysis presented in this paper relies on $\tauhadvis{+}\met$ triggers.  
To correct for any difference between the trigger efficiencies observed in simulation
and those observed in data, $\pt ^{\tau}$-  and $\met$-dependent correction factors are derived,
whose evaluation is limited by statistical uncertainties.  To increase the
sample size, the $\tauhadvis$ and $\met $ trigger efficiencies
are determined separately and residual effects due to correlations are taken into
account as systematic uncertainties. To measure the efficiencies, a tag-and-probe method is used
in a control region enriched with $\ttbar$ events with a $\mu+\tauhad$ selection
using a muon trigger with a $\pt$ threshold of $24$\,GeV\ or $36$\,GeV.
The trigger efficiencies are fitted separately for events with a $\tauhadvis$ that has 
one or three charged-particle tracks. The $\pt^\tau$ ($\met$) trigger efficiencies are fitted in the range of 
20--100 (20--500)\,GeV. The ratios of the fitted functions for data and simulation are then
applied to the simulated samples as continuous correction factors.

Since no trigger information is available in the embedded sample, trigger efficiencies are
applied to that sample. The efficiencies for the $\tauhadvis$ trigger derived as described above need to 
be corrected for misidentified $\tauhadvis$. The fraction of events with a misidentified $\tauhadvis$
is substantial in the $\mu+\tauhad$ sample used for the tag-and-probe method, leading
to a lower efficiency than in a sample with only events that have a true $\tauhadvis$. Since
only events with a true $\tauhadvis$ are present in the embedded sample, the efficiencies determined
from data are corrected by the ratio of the simulated efficiency for true $\tauhadvis$ to the simulated 
efficiency for the $\mu+\tauhad$ sample.

\subsection{Event yields after the event selection}

The expected numbers of 
background events and the 
results from data, together with an expectation from signal
contributions in the low-mass and high-mass $H^+$ selections, are shown in table~\ref{tab:results_taujets_postfit}.  
For the low-mass $H^+$ search, the signal contribution is shown for
a cross section corresponding to ${\cal B}(t \to bH^{+})\times{\cal B}(H^+\rightarrow\tau\nu)=0.9\%$,
and for the high-mass $H^+$ search a possible signal contribution in the \mhmax\ scenario of the MSSM with
$\tan\beta=50$ is shown.

The number of events with a true $\tauhad$ is derived 
from the number of embedded events and does not depend on the theoretical
cross section of the ${t\bar{t} \rightarrow b\bar{b}W^+W^-}$ process. 
However, this analysis does rely on the 
theoretical inclusive \ttbar\ production cross section 
${\sigma_{t\bar{t}} = 253^{+ 13}_{- 15}}$\,pb~\cite{hathor}
for the estimation of the small background with electrons or 
muons misidentified as $\tauhadvis$. 

\begin{table}[h!]
\begin{center}
\begin{tabular}{l|l|l}
Sample & Low-mass $H^+$ selection & High-mass $H^+$ selection \\
\hline \hline
True $\tauhad$ (embedding method) & $2800\pm60\pm500$ &  $3400\pm60\pm400$\\
Misidentified jet$\,\to\tauhadvis$ & $\phantom{0}490\pm\phantom{0}9\pm\phantom{0}80$ & $\phantom{0}990\pm15\pm160$ \\
Misidentified $e\to\tauhadvis$   & $\phantom{00}15\pm\phantom{0}3\pm\phantom{00}6$ & $\phantom{00}20\pm\phantom{0}2\pm\phantom{00}9$\\
Misidentified $\mu\to\tauhadvis$   & $\phantom{00}18\pm\phantom{0}3\pm\phantom{00}8$ & $\phantom{00}37\pm\phantom{0}5\pm\phantom{00}8$\\
\hline
All SM backgrounds   & $3300\pm60\pm500$ & $4400\pm70\pm500$\\
\hline
Data         &  $3244$          &  $4474$       \\
\hline
$H^+$ ($m_{H^{+}}=130$\,GeV) & $\phantom{0}230\pm10\phantom{0}\pm40$ & \\
$H^+$ ($m_{H^{+}}=250$\,GeV) &    & $\phantom{00}58\pm\phantom{0}1\pm\phantom{00}9$\\
\end{tabular}
\caption{\label{tab:results_taujets_postfit} Expected event yields after all selection criteria
and comparison with \LUMI of data. 
The values shown for the signal correspond to the previously published upper limit on ${{\cal B}(t \to bH^{+})\times{\cal B}(H^+\rightarrow\tau\nu)=0.9\%}$~\cite{Aad:2012rjx}
for the low-mass signal point and $\tan \beta=50$ in the MSSM \mhmax~scenario
for the high-mass signal point.
The predicted yield for the low-mass signal selection assumes a $t\bar{t}$ cross section of $253$\,pb.
Both the statistical 
and systematic uncertainties (section~\ref{sec:systematics}) are shown, in this order.
}
\end{center}
\end{table}

%%%%%%%%%%%%%%%%%%%%%%%%%%%%%%%%%%%%%%%%%%%%%%%%%%%%%%%%%%%%%%%%%%%%%%%%%%%%%%%
% Systematics
%%%%%%%%%%%%%%%%%%%%%%%%%%%%%%%%%%%%%%%%%%%%%%%%%%%%%%%%%%%%%%%%%%%%%%%%%%%%%%%

\section{Systematic uncertainties}
\label{sec:systematics}

\subsection{$\tauhadvis{+}\met$ triggers} 
\label{sec:triggersystematics}
\begin{figure}[h!]
\centering
\subfigure[\label{fig:trigger_systs_a} 1-track $\tauhadvis$, $\tauhadvis$ trigger]{
    \includegraphics[width=0.45\textwidth]{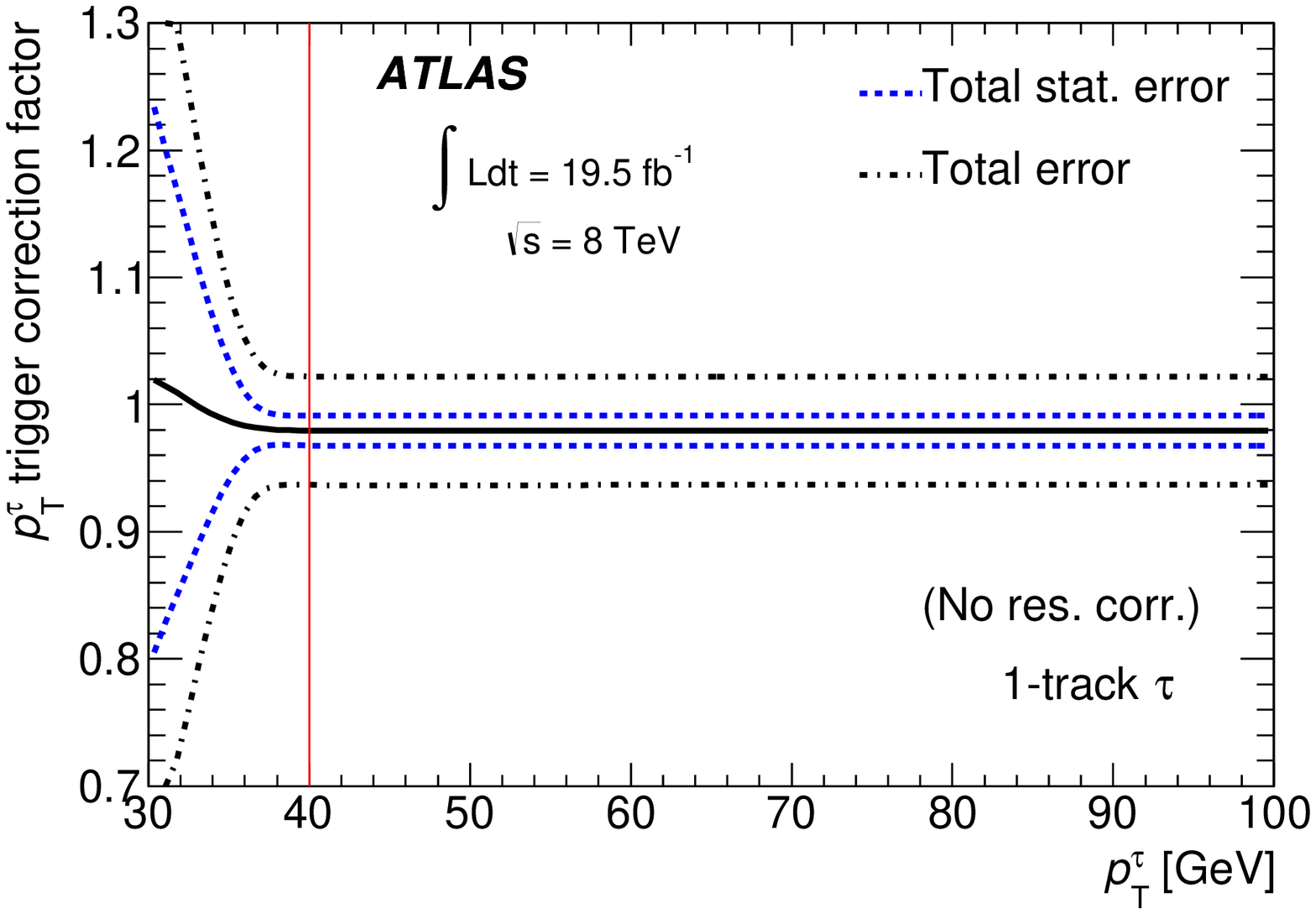}
}
\subfigure[\label{fig:trigger_systs_b} 3-track $\tauhadvis$, $\tauhadvis$ trigger]{
    \includegraphics[width=0.45\textwidth]{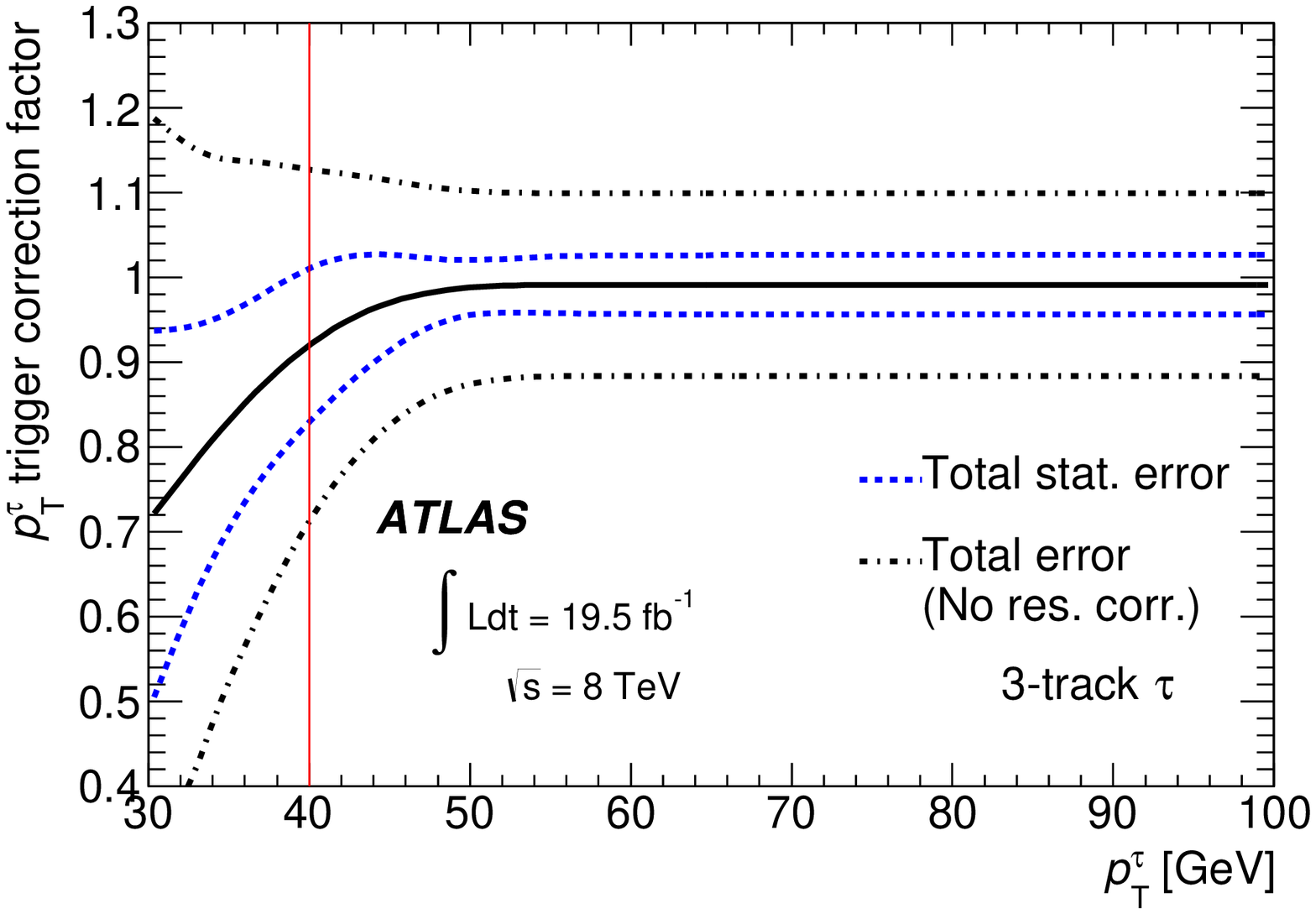}
}
\subfigure[\label{fig:trigger_systs_c} 1-track $\tauhadvis$, $\met$ trigger]{
    \includegraphics[width=0.45\textwidth]{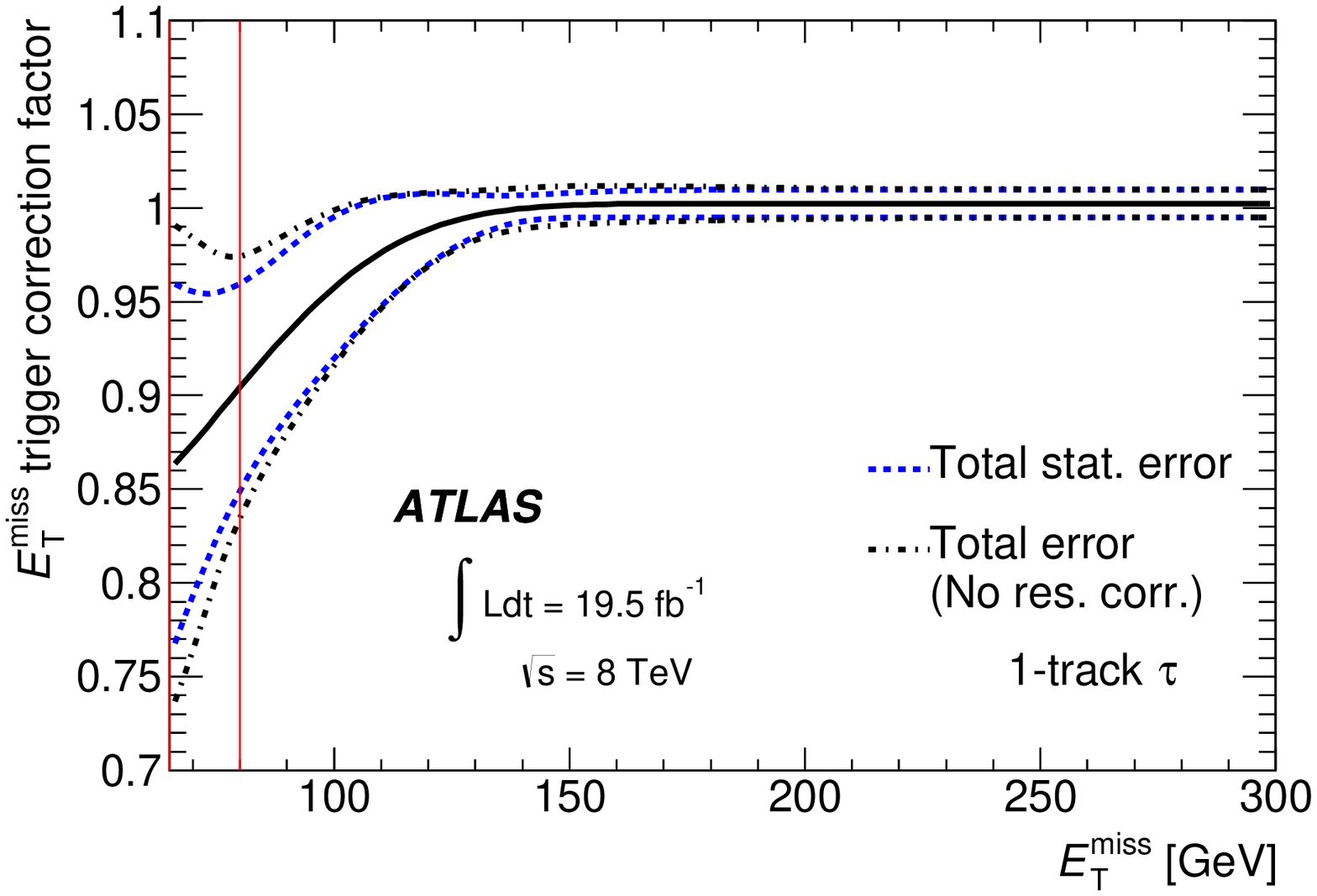}
}
\subfigure[\label{fig:trigger_systs_d} 3-track $\tauhadvis$, $\met$ trigger]{
    \includegraphics[width=0.45\textwidth]{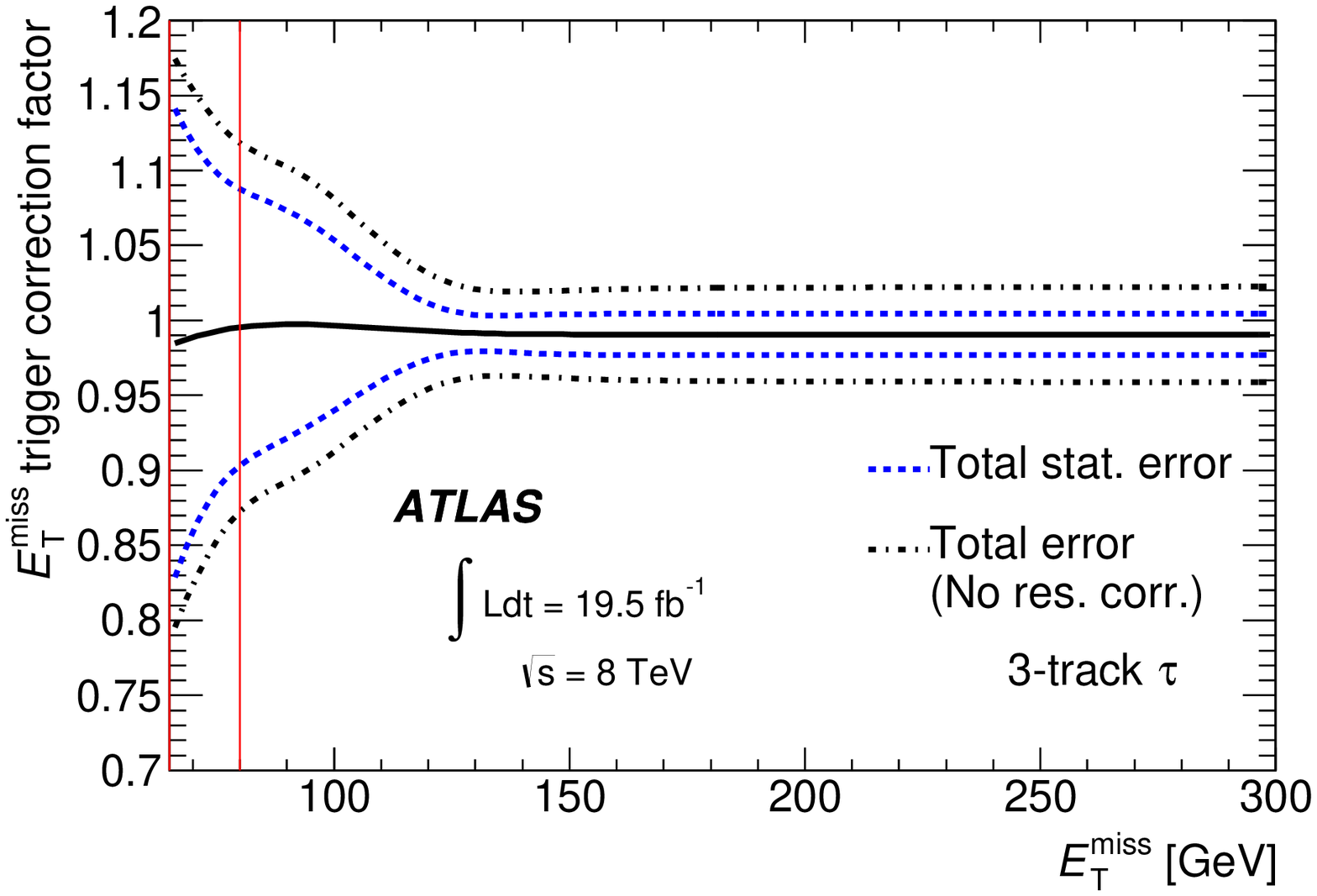}
}
  \caption{Inclusive $\tauhadvis$ and $\met$ trigger correction factors obtained from the ratio of functions fitted to data and simulation 
for $\tauhadvis$ with (a, c) one and (b, d) three charged tracks are shown at the top and bottom, respectively. The vertical line on the $\tauhadvis$ trigger correction factor plots indicates the lowest $\pt^\tau$ 
threshold used in the analysis. The vertical line on the $\met$ trigger correction factor plots shows the lower boundary used in the high-mass
charged Higgs boson search. The total statistical uncertainty is indicated with the dotted line and the systematic uncertainty 
is added in quadrature to the statistical error (dashed-dotted line). For $\tauhadvis$ trigger efficiencies, the total systematic uncertainty is 
shown, while for the $\met$ trigger efficiency, the systematic uncertainty related to residual correlations between $\tauhadvis$ and $\met{}$ is not included, 
since the effect is evaluated separately for several signal mass ranges and background samples.
\label{fig:trigger_systs}}
\end{figure}

Systematic uncertainties on the measurement of the $\tauhadvis{+}\met$ trigger efficiencies
arise from multiple sources: the selection of the muon in the $\mu{+}\tauhad$ sample, the number of
misidentified $\tauhadvis$, the choice of fitting function, slightly varying trigger 
requirements during the data-taking period, a residual correlation between the $\tauhadvis$ and
$\met$ triggers, and the effect of the $\tauhadvis$ energy correction on the trigger efficiency. 
The dominant systematic uncertainty, which arises from misidentified $\tauhadvis$ in the $t\bar{t}\to\mu\tauhad+X$ control region, 
is evaluated by measuring the trigger correction factors after varying the expected misidentified $\tauhadvis$ yield
by its uncertainty.
These uncertainties are relevant for background events with leptons misidentified as $\tauhadvis$
as well as true $\tauhad$ and signal events. The effects on a low-mass and a high-mass signal sample are summarised in table~\ref{tab:triggersysts}
and the effect on background events with true $\tauhad$ is shown in table~\ref{tab:systmethod}.
The trigger correction factors used to account for differences between the efficiencies in
simulation and data are shown in figure~\ref{fig:trigger_systs}.
\begin{table}
\begin{center}
{
\begin{tabular}{l|c|c}
Source of uncertainty & Low-mass $H^+$ selection & High-mass $H^+$ selection \\
\hline\hline

\; Muon selection                  & $ < 1\%$  & $ < 1\%$\\
\; Misidentified $\tauhadvis$      & $ 5.6\%$ & $ 5.7\%$\\
\; Fitting function                & $ 2.1\%$ & $ 1.8\%$\\
\; Trigger definition              & $ < 1\%$  & $ < 1\%$ \\
\; Residual correlations           & $ 1.4\%$ & $ 3.2\%$\\
\; $\tauhadvis$ energy scale             & $ < 1\%$  & $ < 1\%$\\

\end{tabular}
}
\end{center}
\caption{\label{tab:triggersysts}
Effect of systematic uncertainties on the combined trigger efficiencies for a low-mass ($m_{H^+}=130$\,GeV) and high-mass ($m_{H^+}=250$\,GeV) signal sample.
}
\end{table}

\subsection{Data-driven background estimation}
\label{sec:backgroundsystematics}
The systematic uncertainties arising from the data-driven 
methods used to estimate the various backgrounds are summarised 
in table~\ref{tab:systmethod}.\\
\begin{table}
\begin{center}
{
\begin{tabular}{l|c|c}
Source of uncertainty & Low-mass $H^+$ selection & High-mass $H^+$ selection \\
\hline\hline
True $\tauhad$ \\
\hline
\; Embedding parameters           & $3.0\%$  & $1.8\%$ \\
\; Muon isolation                 & $0.3\%$ & $2.3\%$ \\
\; Parameters in normalisation    & $2.0\%$ & $2.0\%$ \\
\; $\tauhadvis$ identification    & $2.2\%$ & $2.0\%$ \\
\; $\tauhadvis$ energy scale      & $4.0\%$ & $3.6\%$ \\
\; $\tauhadvis+\met$ trigger      & $8.3\%$ & $8.3\%$ \\
\hline\hline
Jet $\rightarrow \tauhadvis$ \\
\hline
\; Statistical uncertainty on $p_{\text{m}}$       & $2.0\%$  & $3.4\%$ \\
\; Statistical uncertainty on $p_{\text{r}}$       & $0.5\%$  & $0.5\%$ \\
\; Jet composition               & $1.1\%$  & $1.9\%$ \\
\; $\tauhadvis$ identification         & $0.8\%$  & $0.6\%$ \\
\; $e$/$\mu$ contamination             & $0.5\%$  & $0.7\%$ \\
\end{tabular}
}
\end{center}
\caption{\label{tab:systmethod}
Dominant systematic uncertainties on the data-driven background estimates. The shift in event yield is given
relative to the total background. 
\vspace*{2mm}
}
\end{table}

The systematic uncertainties affecting the 
estimation of the backgrounds 
with true $\tauhad$,  
discussed in section~\ref{sec:embedding_taujets}, consist of the potential 
bias introduced by the embedding method itself (embedding parameters, evaluated by varying the amount of energy
that is subtracted when removing calorimeter deposits of the muon in the original event), uncertainties from 
the trigger efficiency measurement as discussed in section~\ref{sec:triggersystematics},
 uncertainties due to
a possible contamination from multi-jet events (evaluated by varying the muon isolation requirements),
uncertainties associated with the 
simulated $\tauhad$ ($\tauhadvis$ energy scale and identification 
efficiency) and uncertainties on the normalisation. The latter 
are dominated by the statistical uncertainty of the selected control 
sample and the $\tauhadvis{+}\met$ trigger efficiency uncertainties.

For the estimation of backgrounds with jets 
misidentified as $\tauhadvis$, discussed 
in section~\ref{sec:qcd_taujets}, 
the dominant systematic uncertainties on the misidentification 
probability are the statistical uncertainty due to the 
control sample size and uncertainties due to the difference in 
the jet composition (gluon- or quark-initiated) between the control 
and signal regions.  The uncertainty arising from differences in jet
composition is evaluated from the difference in shape and normalisation that
arises when $p_{\text{m}}$ is measured in a control region that is enriched
in events with gluon-initiated jets. This control region differs from the signal 
region only by inverting the $b$-tag and $\met$ requirements. 
Other uncertainties are due to the statistical
uncertainty on $p_{\text{r}}$, 
the effect of the uncertainty in the simulated 
true $\tauhadvis$ identification efficiency on the measurement of $p_{\text{r}}$ and $p_{\text{m}}$, and
the effect of the simulated electron veto efficiency for true electrons on the measurement of $p_{\text{m}}$.

\subsection{Detector simulation}
\label{ref:detectorsystematics}
Systematic uncertainties originating from the simulation of pile-up and object reconstruction  
are considered for signal events and background events with leptons misidentified as $\tauhadvis$. This background 
is roughly 1$\%$ of the final background in both the low-mass and high-mass searches, so the systematic 
uncertainties have little effect on the final results.  

Uncertainties related to the $\tauhadvis$ energy scale and identification efficiency 
are also taken into account. The uncertainty on the 
identification is in the range $2{-}3$\% for $\tauhadvis$ with
one charged track and $3{-}5$\% for $\tauhadvis$ with three charged tracks. It has been
measured in data using a tag-and-probe method~\cite{taus}. The $\tauhadvis$ energy scale is measured
 with a precision of $2{-}4$\%~\cite{ATLAS-CONF-2013-044}. It is determined by fitting 
the reconstructed visible mass of ${Z\rightarrow \tau\tau}$ events in data. Uncertainties 
related to jet or $b$-tagged jet energy scale, energy resolution, flavour identification and calibration, and 
the effects of pile-up interactions are also taken into account \cite{Aad:2014bia,b_eff}, 
as well as uncertainties related to the reconstruction of $\met$ \cite{Aad:2012re}. 

The impact of most sources of systematic uncertainty is
assessed by re-applying the selection cuts for each analysis after
varying a particular parameter by its $\pm1$ standard deviation uncertainty.
The dominant instrumental systematic uncertainties include the jet energy scale,
the $\tauhadvis$ energy scale, and $\tauhadvis$ identification. All instrumental systematic 
uncertainties are taken into account for the reconstruction of $\met$.

These uncertainties affect all simulated samples, i.e.\ the signal and the background contribution
with leptons misidentified as $\tauhadvis$. Since the $\tauhad$ in the embedded samples are
simulated, all $\tauhad$-related uncertainties are relevant for the background with true $\tauhad$ as well.

\subsection{Generation of \ttbar\ and signal events}
\label{sec:generatorsystematics}
In order to estimate the systematic uncertainties arising from the
$t\bar t$ and low-mass signal generation, as well as from the parton shower model, the acceptance
is computed for $t\bar t$ events produced with MC@NLO
interfaced to HERWIG/JIMMY and POWHEG 
interfaced to PYTHIA 8. Also, an uncertainty on the
theoretical cross section, including both the factorisation/renormalisation scale
and parton distribution function uncertainties, is taken into account for $\ttbar$ 
backgrounds with a lepton misidentified as a $\tauhadvis$ and low-mass signal samples.  
The estimate of the small background with electrons or
muons misidentified as $\tauhadvis$ relies additionally on the theoretical inclusive \ttbar\ production cross section
${\sigma_{t\bar{t}} = 253^{+ 13}_{- 15}}$\,pb~\cite{hathor}.

The generator modelling uncertainties for the high-mass signal samples are
estimated from a comparison between events produced with MC@NLO interfaced to
HERWIG++~\cite{Herwigpp} and POWHEG interfaced to PYTHIA 8.  

The systematic uncertainties originating from
initial- and final-state parton radiation, which modify the jet production rate, are computed for 
$\ttbar$ backgrounds and applied to low-mass signal events by using $\ttbar$ samples generated with AcerMC 
interfaced to PYTHIA 6, where initial- and final-state
radiation parameters are set to a range of values not excluded by the 
experimental data~\cite{ATL-PHYS-PUB-2014-005}. The largest relative differences with 
respect to the reference sample, after full event selections, are used as 
systematic uncertainties. For high-mass signal samples, this uncertainty is evaluated 
by varying factorisation/renormalisation scale parameters in the production of signal samples (QCD scale). The uncertainty due   
to the choice of parton distribution function has a negligible impact for both background and signal, and is not included. An additional 
uncertainty, arising from the difference in acceptance between 4FS and 5FS $H^+$ production
is evaluated using dedicated signal samples that are generated at leading order with MadGraph~\cite{Alwall:2011uj} interfaced with PYTHIA~8, although the 
nominal signal samples are generated at NLO. The systematic uncertainties arising from the modelling of the $t\bar{t}$  and signal
event generation and the parton shower, as well as from the initial- and final-state 
radiation, are summarised in table~\ref{tab:sysgen}.

All of these uncertainties, except for $H^+$ production, affect only signal and background events where leptons are misidentified as $\tauhadvis$.
\begin{table}[h!]
  \begin{center}
    \begin{tabular}{l|c}
        Source of uncertainty & Normalisation uncertainty\\
      \hline \hline
      Low-mass $H^{+}$ & \\
       \hline
      Generator model ($\bbbar W^{-}H^{+}$) & $\phantom{0{-}{}}9\%$ \\
      Generator model ($\bbbar W^{+}W^{-}$) & $\phantom{0{-}{}}9\%$ \\
      $\ttbar$ cross section & $\phantom{0{-}{}}6 \%$\\
      Jet production rate (SM and $H^{+}$) (QCD scale)& $\phantom{{}{-}{}}11\%$ \\
\hline\hline
      High-mass $H^{+}$ & \\
       \hline
      Generator model ($H^{+}$) & $2{-}9\%$ \\
      Generator model (SM) & $\phantom{0{-}{}}8\%$ \\
      $\ttbar$ cross section & $\phantom{0{-}{}}6 \%$\\
      Jet production rate ($H^{+}$) (QCD scale) &$1{-}2\%$ \\
      Jet production rate (SM) (QCD scale)& $\phantom{{}{-}{}}11\%$ \\
      $H^+$ production (4FS vs 5FS)& $3{-}5\%$ \\
    \end{tabular}
    \caption{Systematic uncertainties arising from $\ttbar$ and signal generator modelling, and from the jet production rate.
             The uncertainties are shown for the $\ttbar$ background and the charged Higgs boson signal, for the low-mass and high-mass charged
             Higgs boson selections separately.  The systematic uncertainty of the $H^{+}$ yield due to QCD scale and 4FS vs 5FS production
             was evaluated at masses of 200, 400, and 600 GeV.  For all other systematic variations of the $H^{+}$ yield, all mass
             points were considered.
      \label{tab:sysgen}}
  \end{center}
\end{table}

%%%%%%%%%%%%%%%%%%%%%%%%%%%%%%%%%%%%%%%%%%%%%%%%%%%%%%%%%%%%%%%%%%%%%%%%%%%%%%%
% Limits
%%%%%%%%%%%%%%%%%%%%%%%%%%%%%%%%%%%%%%%%%%%%%%%%%%%%%%%%%%%%%%%%%%%%%%%%%%%%%%%

\section{Statistical analysis}
\label{sec:limit_setting}

In order to test the compatibility of the data with background-only 
and signal+background hypotheses, a profile log-likelihood 
ratio~\cite{PhysRevD.86.032003} is used with $m_{\text{T}}$
as the discriminating variable. The statistical analysis is based on a 
binned likelihood function for these distributions. Systematic 
uncertainties 
in shape and normalisation, discussed in section~\ref{sec:systematics}, are incorporated via nuisance parameters fully
correlated amongst the different backgrounds, and 
the one-sided profile likelihood ratio, $\tilde{q}_{\mu}$, is used as a 
test statistic. The parameter of interest, the signal-strength $\mu$, is either 
${{\cal B}(t\rightarrow bH^+)\times {\cal B}(H^+\rightarrow\taup\nu)}$ (low-mass search) or $\xsection \times \brhtn$ (high-mass search).
Expected limits are derived 
using the asymptotic approximation~\cite{Cowan:2010js}.

\begin{figure}[!ht]
  \centering
\subfigure[\label{fig:pulls_lowmass} Low-mass $H^+$ selection]{
\includegraphics[width=0.47\textwidth]{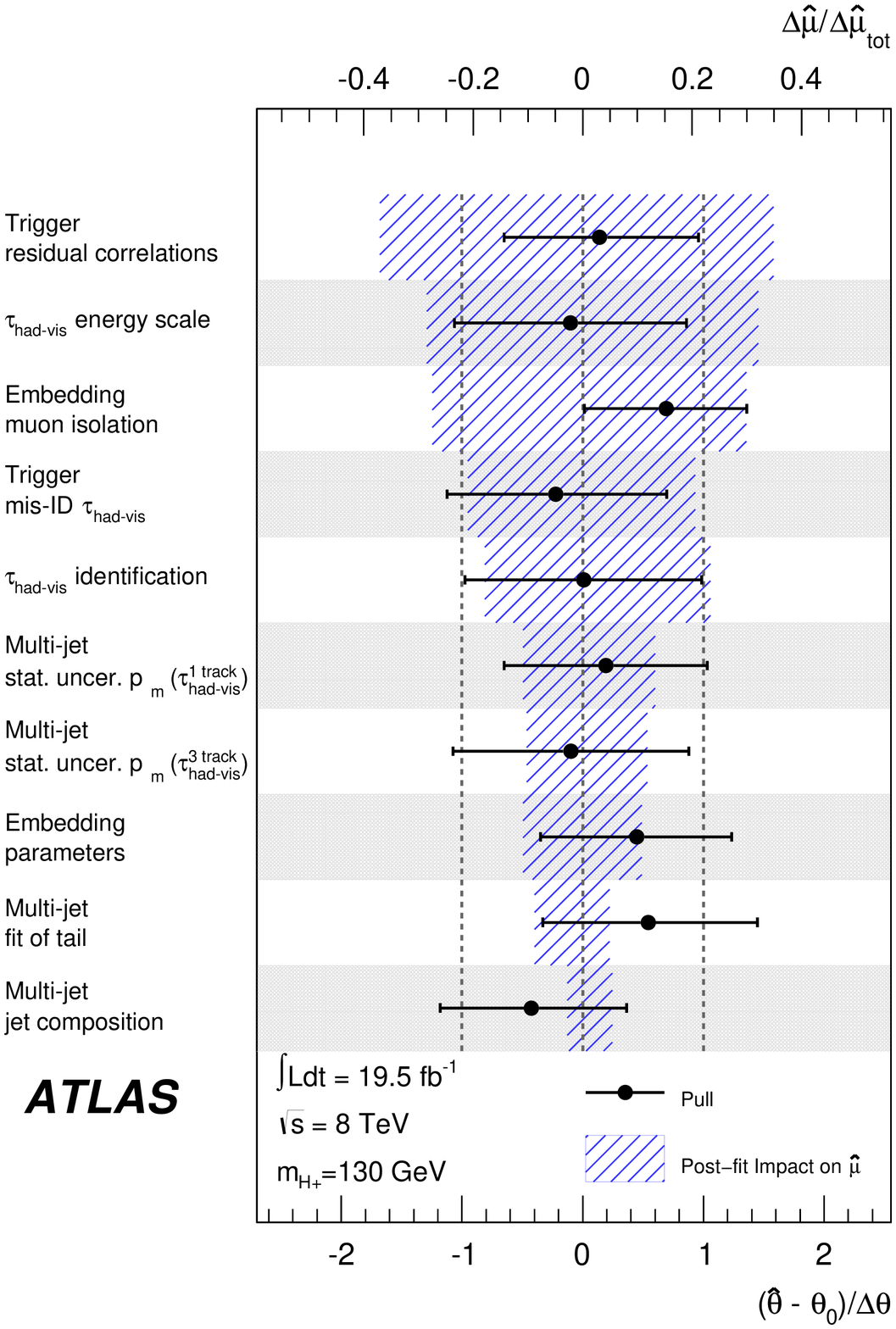}}
\subfigure[\label{fig:pulls_highmass} High-mass $H^+$ selection]{
\includegraphics[width=0.47\textwidth]{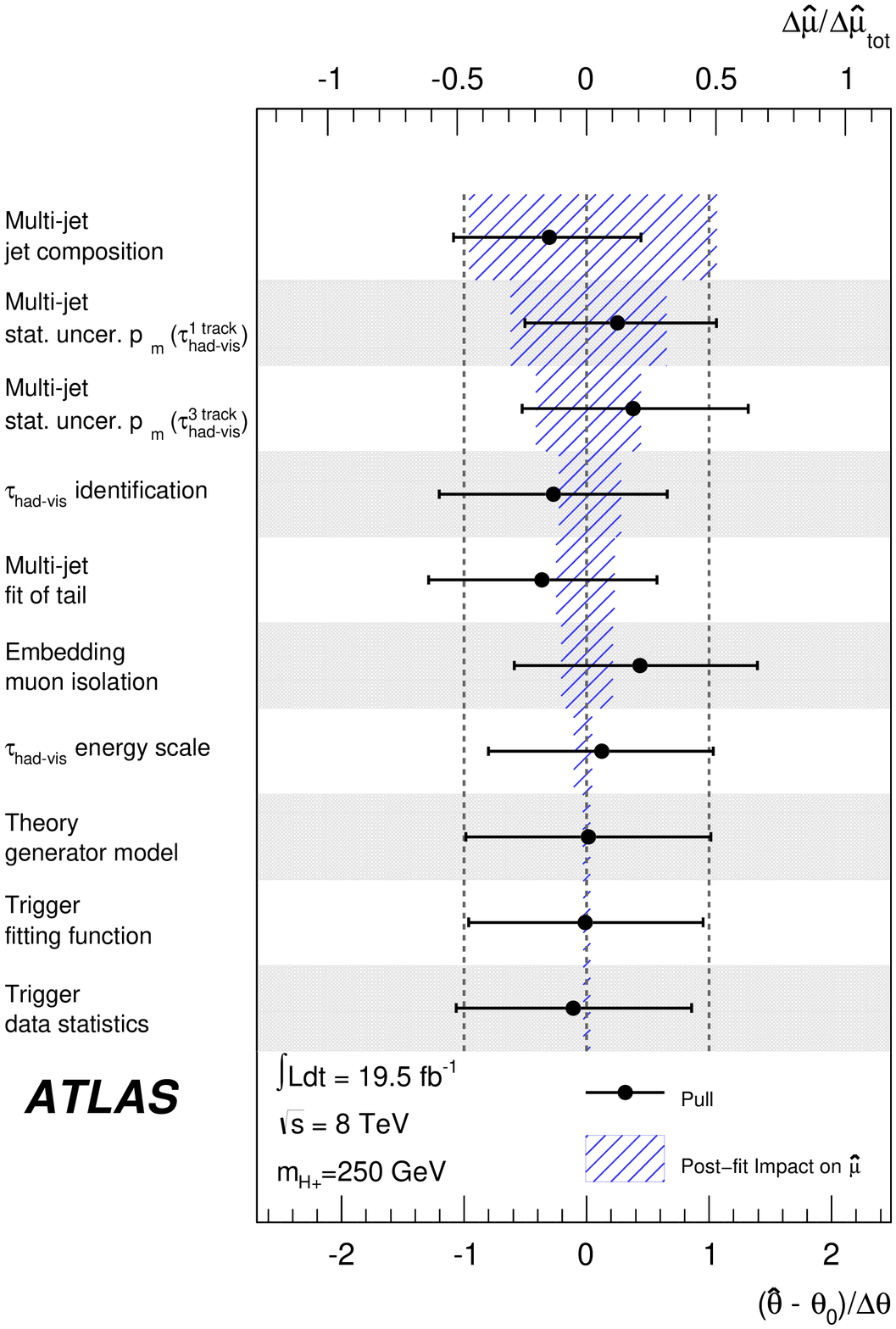}}
\caption{\label{fig:pulls} Impact of systematic uncertainties on the final
observed limits for (a) $m_{H^+}=130$\,GeV\ and (b) $m_{H^+}=250$\,GeV.
The systematic uncertainties are ordered (top to bottom) by decreasing impact on the fitted signal
strength parameter. 
The dots, which refer to the bottom horizontal axis, show 
how each fitted nuisance parameter, $\hat{\theta}$, deviates from its nominal
value, $\theta_{0}$, in terms of standard deviations with respect to its
nominal uncertainty, $\Delta\theta$. 
The solid lines indicate the post-fit uncertainties of each nuisance parameter,
also relative to their nominal values.
The hatched band, referring to the top horizontal axis,
shows the deviations of the fitted signal-strength
parameter after changing a specific nuisance parameter
upwards or downwards by its post-fit uncertainty ($\Delta\hat{\mu}$) as a fraction of the total uncertainty
of the fitted signal-strength parameter~($\Delta\hat{\mu}_\mathrm{tot}$).
}
\end{figure}

The nuisance parameters are simultaneously fitted by means of a negative log-likelihood minimisation in both low-mass and high-mass regions 
in order to ensure that they are well estimated.  
This is shown in figure~\ref{fig:pulls} for the nuisance parameters that have the largest impact on the fitted $\mu$, denoted $\hat{\mu}$. 
The black dots indicate how a given nuisance parameter deviates from expectation, while the
black error bars indicate how its post-fit uncertainty compares with its nominal uncertainty. In both
the low-mass and high-mass searches, the black dots and error bars indicate respectively that none of the nuisance parameters deviate
by more than one standard deviation and that their uncertainties are not underestimated. The blue
hatched box shows the deviations of the fitted signal-strength
parameter after changing a specific nuisance parameter
upwards or downwards by its post-fit uncertainty.

The results in figure~\ref{fig:pulls} indicate the relative impact of the
systematic uncertainties in the statistical analysis of the low-mass and high-mass searches.
For the low-mass search, the most important systematic uncertainties are
those related to the measurement of the trigger efficiency and to the
simulation of the detector response to $\tauhadvis$. Since
the low-mass search is dominated by the presence of backgrounds with 
a true $\tauhad$, this is consistent with expectations.
For the high-mass search, the most important systematic uncertainties
are due to jets misidentified as $\tauhadvis$,
including both the yield and $m_{\text{T}}$ distribution of such events, and 
the next dominant effect is from the true $\tauhadvis$ background.

\section{Results}
\label{sec:limit}

\begin{figure}[h!]
\begin{center}
\subfigure[\label{fig:mT_lowmass} Low-mass $H^+$ selection]{
  \includegraphics[width=0.47\textwidth]{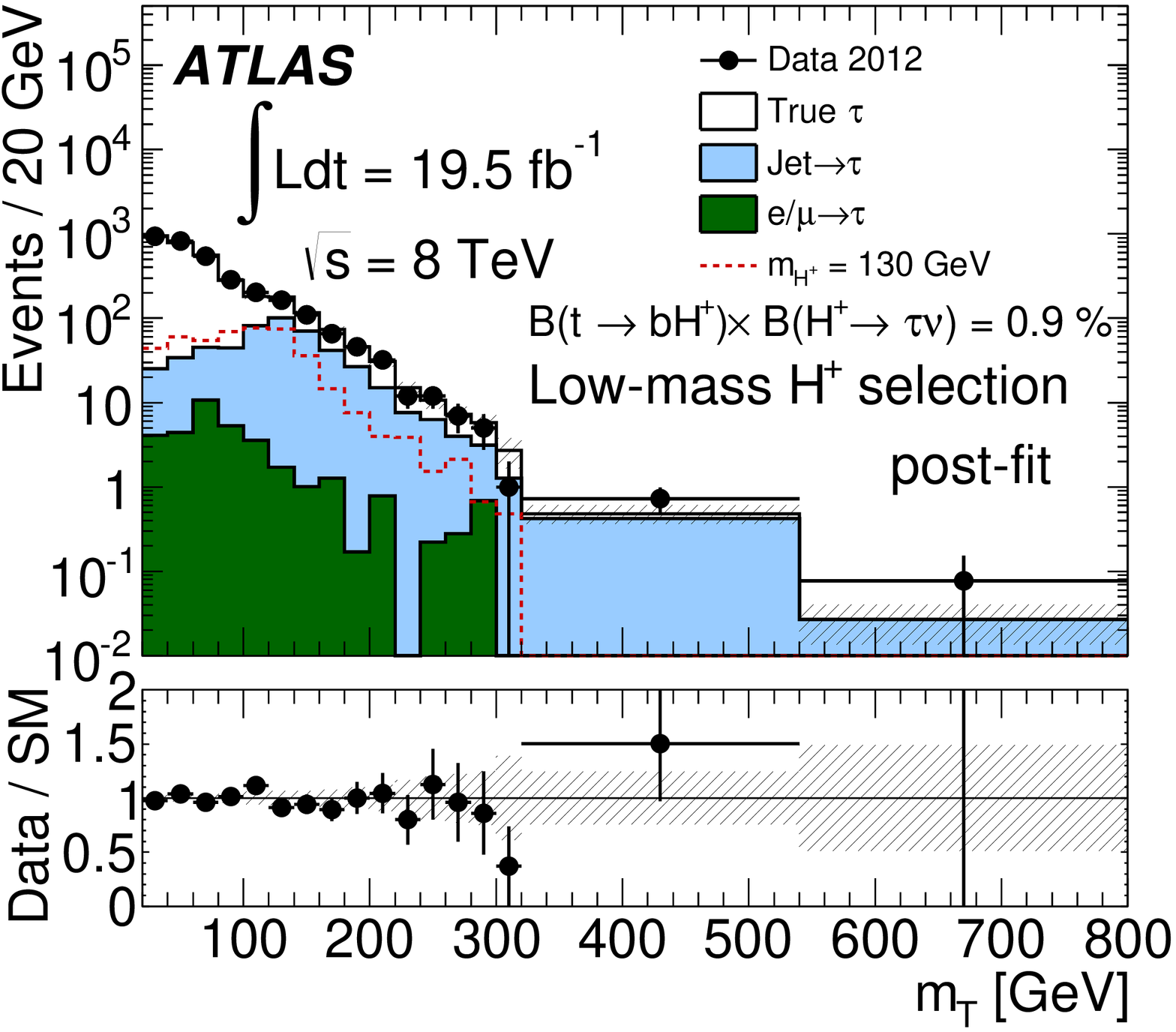}
}
\subfigure[\label{fig:mT_highmass} High-mass $H^+$ selection]{
  \includegraphics[width=0.47\textwidth]{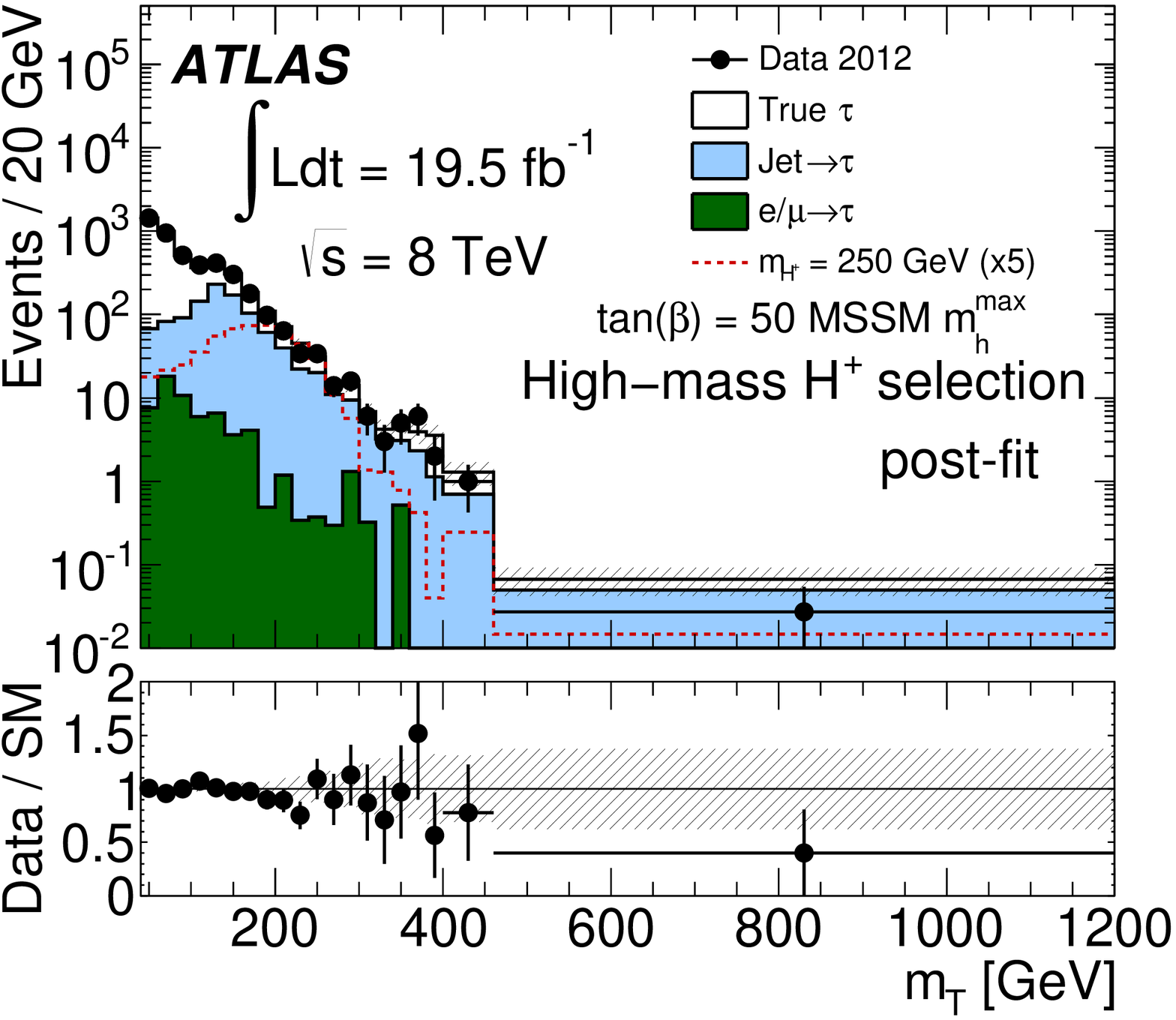}
}
\end{center}
\caption{
Distributions of $m_{\text{T}}$ after all selection criteria. The hatched area shows
the total post-fit uncertainty for the SM backgrounds. For the (a) low-mass selection, bins
are 20\,GeV\ wide up to $m_{\text{T}}$ = 320\,GeV, then 320--540\,GeV\ and $>$ 540\,GeV. For the (b) high-mass selection, 
bins are 20\,GeV\ wide up to $m_{\text{T}}$ = 400\,GeV, then 400--460\,GeV\ and $>$ 460\,GeV.
All bins are normalised to a 20\,GeV\ bin width.
For the low-mass search (a),
a possible signal contribution with $m_{H^+} = 130~\mbox{\GeV}$,
and ${{\cal B}(t \to bH^{+})\times{\cal B}(H^+\rightarrow\taup\nu)=0.9\%}$ is overlaid on top of the SM contributions.
For the high-mass search (b), a possible signal contribution with $m_{H^+} = 250~\mbox{\GeV}$ and $\tan{\beta} = 50$
in the \mhmax\ scenario of the MSSM, where the corresponding cross section~\cite{Flechl:2014wfa} is scaled up by a factor
of five, is overlaid on the SM contributions. 
\label{fig:results_taujets_postfit}}
\end{figure}

\begin{figure}[!ht]
  \centering
\subfigure[\label{fig:limit_lowmass} Low-mass $H^+$ selection]{
\includegraphics[width=0.47\textwidth]{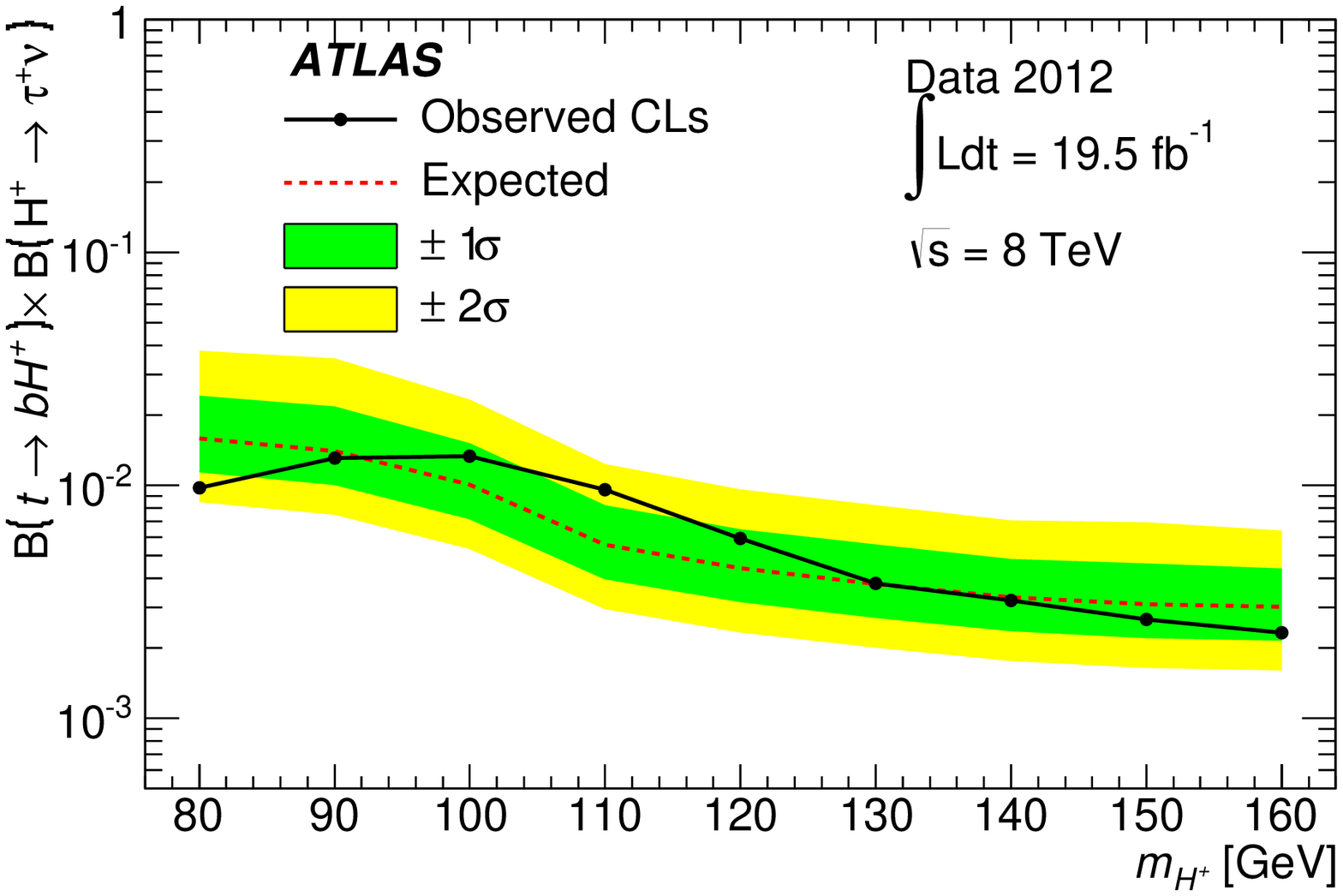}}
\subfigure[\label{fig:limit_highmass} High-mass $H^+$ selection]{
\includegraphics[width=0.47\textwidth]{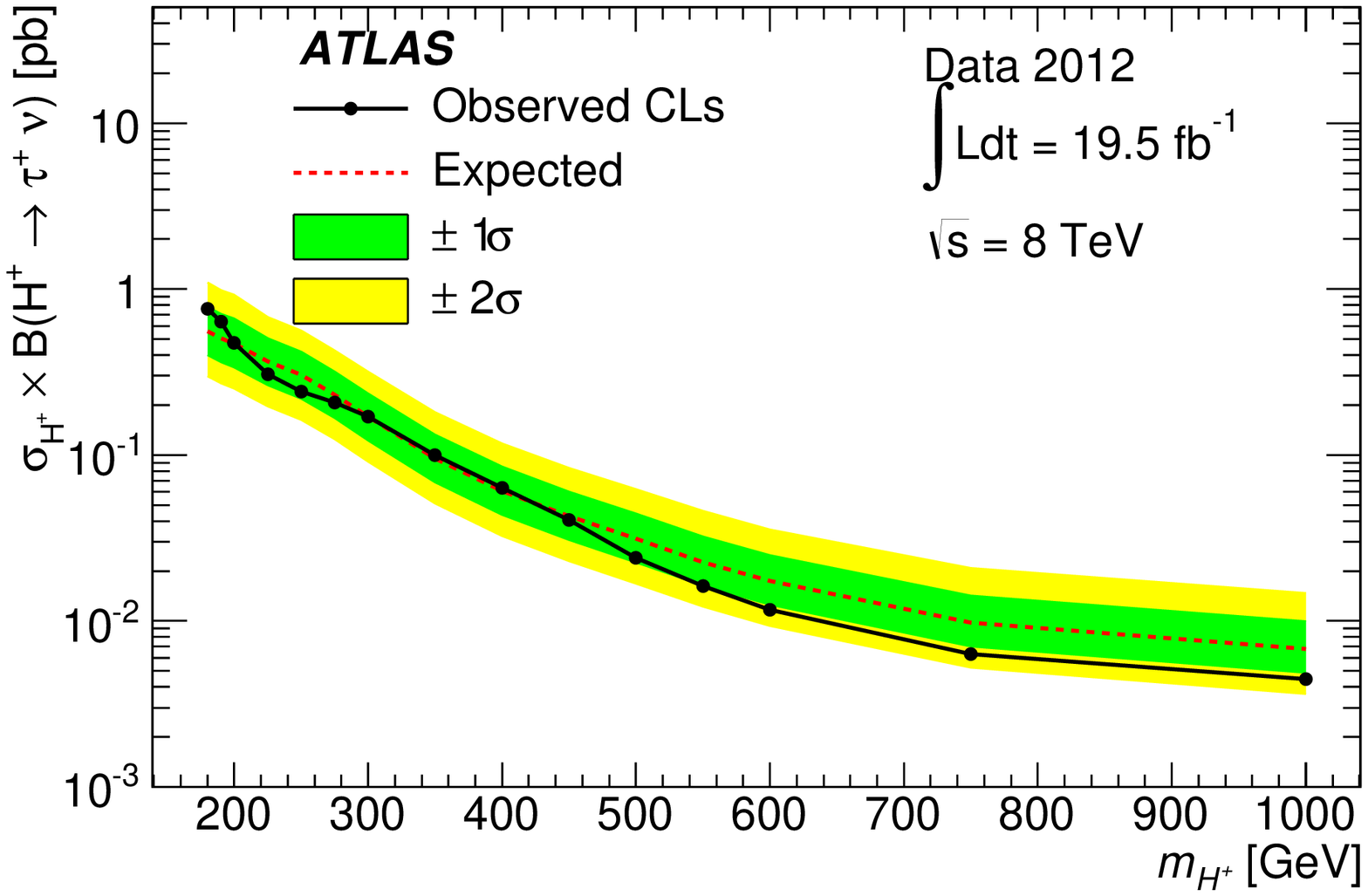}}
\caption{\label{fig:limit} Observed and expected
95$\%$ CL exclusion limits on
the production and decay of (a) low-mass and (b) high-mass charged Higgs
bosons. For the low-mass search, the limit is computed for 
${{\cal B}(t\rightarrow bH^+)\times{\cal B}(H^+ \to \taup\nu)}$ for charged Higgs boson production from top-quark decays as a function of $m_{H^+}$.
For the high-mass search,
the limit is computed for $\xsection \times  \brhtn$, and is to be understood as applying to 
the total production cross section times branching ratio of $H^+$ and $H^-$ combined.
}
\end{figure}

In figure~\ref{fig:results_taujets_postfit}, the \mt\ distribution after the final fit is shown.
No significant deviation of the data from the SM prediction is observed. 
For the low-mass charged Higgs boson search, exclusion limits are set on the 
branching ratio ${\cal B}(t \to bH^+)\times{\cal B}(H^+ \to \taup\nu)$.  
For the high-mass $H^+$ search, exclusion limits are set on 
$\xsection \times \brhtn$, and are to be understood as applying to the total production
cross section times branching ratio of $H^+$ and $H^-$ combined.
Using the binned log-likelihood described in section~\ref{sec:limit_setting}, all exclusion limits are set by rejecting 
the signal hypothesis at the 95\% confidence level (CL) using the CL$_s$ 
procedure~\cite{Read:2002hq}. These limits are based on the 
asymptotic distribution of the test statistic~\cite{Cowan:2010js}. 
The exclusion limits are shown in figure~\ref{fig:limit_lowmass} for the low-mass search and in figure~\ref{fig:limit_highmass} for the high-mass search.
Expected and observed limits agree well and are within the uncertainties over the whole investigated mass range.
The limits are in the range between 1.3\% and 0.23\% for the low-mass search. 
For the high-mass search, they range from 0.76\,pb to 4.5\,fb in the mass range 180\,GeV\ $\leq m_{H^+}\leq 1000$\,GeV.

The limits on ${{\cal B}(t\rightarrow bH^+)\times{\cal B}(H^+ \to \taup\nu)}$ for the low-mass search and on
$\xsection \times \brhtn$ 
for the high-mass search 
are also interpreted in the context of different scenarios of the MSSM~\cite{Carena:2013qia}.
In the \mhmax\ scenario, the mass of the light CP-even Higgs boson $h$ ($m_h$) is maximised.  
Interpreting the Higgs boson discovered at the LHC as the $h$, only a small region of 
the $m_{H^+}-\tan\beta$ parameter space in this scenario is compatible with the observation.
The \mhmodp\ and \mhmodm scenarios are modifications of the \mhmax\ scenario. The discovered
Higgs boson is interpreted as the $h$ as well but the requirement that $m_h$ be maximal
is dropped. This is done by reducing the amount of
mixing in the top squark sector compared to the \mhmax\ scenario, leading to a larger region
in the parameter space being compatible with the observation.
The \mhmodp\ and \mhmodm\ scenarios only differ in the sign of a parameter.

\begin{figure}[t!h!]
  \centering
\subfigure[\label{fig:mhmax_lowmass} \mhmax\ scenario, low-mass $H^+$ selection]{    
\includegraphics[width=0.4\textwidth]{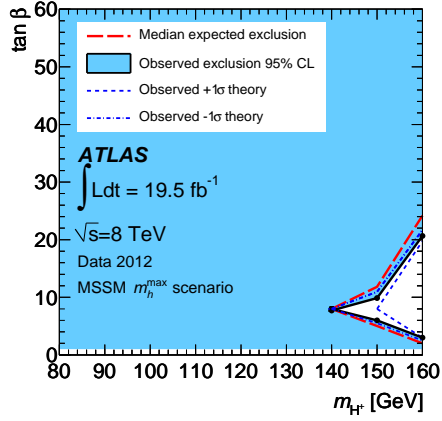}
}
\subfigure[\label{fig:mhmax_highmass} \mhmax\ scenario, high-mass $H^+$ selection]{    
    \includegraphics[width=0.4\textwidth]{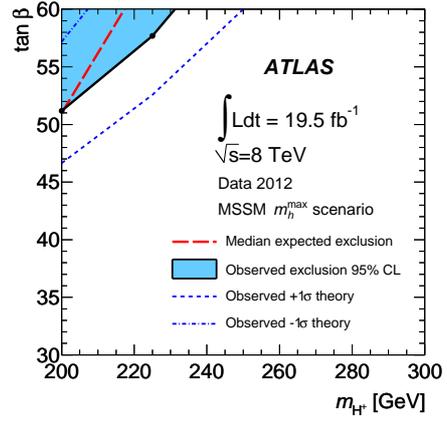}
}
\subfigure[\label{fig:mhmodp_lowmass} \mhmodp\ scenario, low-mass $H^+$ selection]{    
    \includegraphics[width=0.4\textwidth]{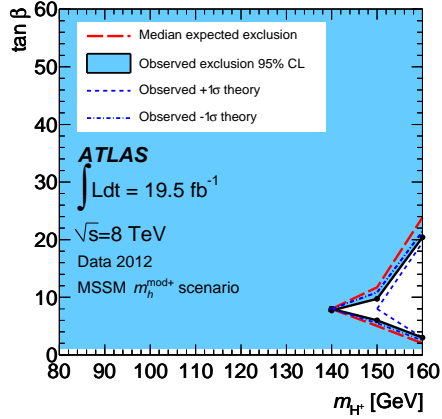}
}
\subfigure[\label{fig:mhmodp_highmass} \mhmodp\ scenario, high-mass $H^+$ selection]{    
    \includegraphics[width=0.4\textwidth]{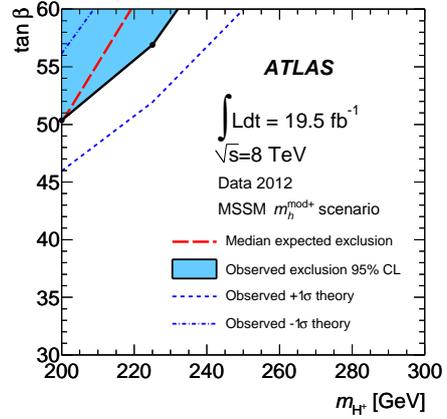}
}
 \subfigure[\label{fig:mhmodm_lowmass} \mhmodm\ scenario, low-mass $H^+$ selection]{    
   \includegraphics[width=0.4\textwidth]{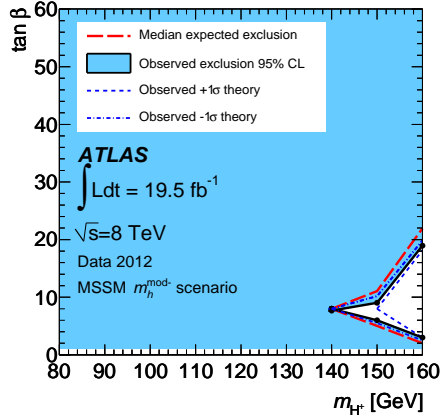}
}
\subfigure[\label{fig:mhmodm_highmass} \mhmodm\ scenario, high-mass $H^+$ selection]{     
   \includegraphics[width=0.4\textwidth]{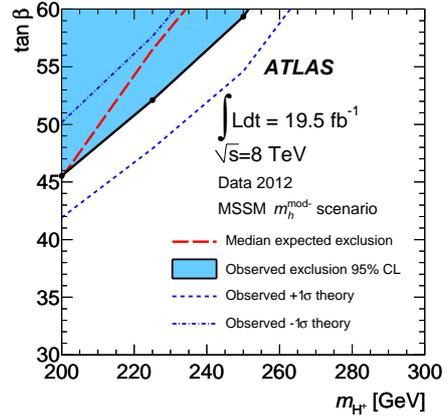}
}
    \caption{The 95\% CL exclusion limits on $\tan\beta$ as a function of 
$m_{H^+}$. Results are shown in the context of different benchmark scenarios of the MSSM 
for the regions in which reliable theoretical predictions exist. Results are shown for (low-mass, high-mass) 
$H^+$ search in the (a, b) \mhmax, (c, d) \mhmodp\ and (e, f) \mhmodm\ scenarios in the left (right) column. 
\label{fig:mssm_limit}}
\end{figure}

Interpretations of the 95\% CL limits in the \mhmax, \mhmodp\ and \mhmodm\ scenarios are shown in figure~\ref{fig:mssm_limit}.
In the low-mass range, almost all values for $\tan \beta >1$ are excluded in the different scenarios, except for a small region ${140\text{\,GeV}\leq m_{H^+}\leq 160}$\,GeV.
Values of $\tan\beta$ larger than $45-50$ are excluded in a mass range of 200\,GeV $ \leq m_{H+} \leq $ 250\,GeV. % for the the \mhmax, \mhmodp~ and \mhmodm~ scenarios.
The exclusions in several additional scenarios, not shown here, were also
considered.
In the light top squark, light stau and tauphobic scenarios, no significant exclusion is achieved in the high-mass search.
In the low-mass search, the excluded regions in these scenarios are similar to those shown in figure~\ref{fig:mssm_limit}. 
The limits for the low-mass $H^+$ search are also interpreted in the low-$M_H$ scenario, where ${m_{H^+}\approx 130}$~\,GeV.
Instead of excluding areas in the $m_{H^+}$--$\tan\beta$ plane, limits
are interpreted in the $\tan\beta$--$\mu$ plane,
for 300\,GeV $< \mu <$ 3500\,GeV and ${1.5<\tan\beta<9.5}$,
where $\mu$ is the higgsino mass parameter. 
This model is excluded everywhere where it is tested and where it is well-defined.
For the interpretation of the low-mass search, the following relative theoretical uncertainties on 
${{\cal B}(t \to bH^+)\times{\cal B}(H^+ \to \taup\nu)}$ are considered~\cite{Heinemeyer:1998yj,Dittmaier:2012vm,PhysRevD.83.055005}: 
5\% for one-loop electroweak corrections missing from the calculations, 
2\% for missing two-loop QCD corrections and about 1\% (depending on 
$\tan \beta$) for $\Delta_b$-induced uncertainties, where $\Delta_b$ is a 
correction factor to the running $b$-quark mass~\cite{Carena:1999py}. 
These uncertainties are added linearly, as recommended by 
the LHC Higgs cross section working group~\cite{Dittmaier:2012vm}. 
For the interpretation of the high-mass search, %the production
separate uncertainties are included for the 4FS and 
5FS calculations \cite{Heinemeyer:2013tqa}. For the 5FS calculation, the following theoretical uncertainties are taken into account: scale uncertainties of
approximately 10--20$\%$ that vary with $m_{H^+}$, the combined uncertainty on the parton distribution function, mass of the $b$-quark,
and strong coupling of approximately 10--15$\%$. For the 4FS calculation, only a scale uncertainty
of approximately 30$\%$ is taken into account.
Owing to the complication arising from the overlap and interference with off-shell $\ttbar$ production in
the mass range of $m_{H^+}=180$--$200$\,GeV, the MSSM interpretation is shown only for $m_{H^+} \geq $ 200\,GeV. 

%%%%%%%%%%%%%%%%%%%%%%%%%%%%%%%%%%%%%%%%%%%%%%%%%%%%%%%%%%%%%%%%%%%%%%%%%%%%%%%
% Conclusion
%%%%%%%%%%%%%%%%%%%%%%%%%%%%%%%%%%%%%%%%%%%%%%%%%%%%%%%%%%%%%%%%%%%%%%%%%%%%%%%

\FloatBarrier
\section{Conclusions}
\label{sec:conclusion}

Charged Higgs bosons decaying via ${H^+\rightarrow\taup\nu}$ are searched 
for in \ttbar\ events, in the decay 
mode $t \to bH^+$  (low-mass search), and for $H^+$ production in association with a top quark, $pp \to tH^{+}+X$ 
(high-mass search). The analysis makes use 
of a total of \LUMI of $pp$ collision data at $\sqrt{s}=8$\,TeV, recorded in 
2012 with the ATLAS detector at the LHC. The final state considered in
this search is characterised by the presence 
of a hadronic $\tau$ decay, missing transverse momentum, $b$-tagged jets, a  
hadronically decaying $W$ boson, as well as the absence of any electrons or muons. Data-driven methods are employed to 
estimate the dominant background contributions.
The data are found to be in agreement with the SM predictions. 
Upper limits at the 95\% confidence level are set on the 
branching ratio ${\cal B}(t \to bH^{+})\times{\cal B}(H^+\rightarrow\taup\nu)$ between 0.23\% and 1.3\% 
for a mass range of $m_{H^+}=$ 80--160\,GeV, a major improvement over the previous published limits of 
0.8--3.4$\%$ for the mass range of $m_{H^+}=$ 90--160\,GeV~\cite{Aad:2012rjx,Chatrchyan:2012vca}.  For the mass range of 
$m_{H^+}=$ 180--1000\,GeV, the first upper limits from ATLAS are set for the 
production cross section times branching ratio, 
$\xsection \times \brhtn$, between 0.76\,pb 
and 4.5\,fb. 
Interpreted in the context of the \mhmax, \mhmodp\ and \mhmodm\ scenarios of the MSSM, the entire parameter space 
with $\tan \beta > 1$ is excluded for the low-mass range 90\,GeV\ $\leq m_{H^+}\leq 140$\,GeV, and almost all of the 
parameter space with $\tan \beta > 1$ is excluded for 140\,GeV\ $\leq m_{H^+}\leq 160$\,GeV.
For the high-mass range 200\,GeV\ $ \leq m_{H^+} \leq 250$\,GeV, a region of parameter space with high $\tan\beta$ is excluded.

%%%%%%%%%%%%%%%%%%%%%%%%%%%%%%%%%%%%%%%%%%%%%%%%%%%%%%%%%%%%%%%%%%%%%%%%%%%%%%%
% Bibliography
%%%%%%%%%%%%%%%%%%%%%%%%%%%%%%%%%%%%%%%%%%%%%%%%%%%%%%%%%%%%%%%%%%%%%%%%%%%%%%

%% \documentclass[11pt,a4paper,dvips]{article}

%% \begin{document}

% Acknowledgements for papers with collision data
% Version 01-Mar-2014

\section{Acknowledgements}

% Standard acknowledgements start here
%----------------------------------------------
We thank CERN for the very successful operation of the LHC, as well as the
support staff from our institutions without whom ATLAS could not be
operated efficiently.

We acknowledge the support of ANPCyT, Argentina; YerPhI, Armenia; ARC,
Australia; BMWFW and FWF, Austria; ANAS, Azerbaijan; SSTC, Belarus; CNPq and FAPESP,
Brazil; NSERC, NRC and CFI, Canada; CERN; CONICYT, Chile; CAS, MOST and NSFC,
China; COLCIENCIAS, Colombia; MSMT CR, MPO CR and VSC CR, Czech Republic;
DNRF, DNSRC and Lundbeck Foundation, Denmark; EPLANET, ERC and NSRF, European Union;
IN2P3-CNRS, CEA-DSM/IRFU, France; GNSF, Georgia; BMBF, DFG, HGF, MPG and AvH
Foundation, Germany; GSRT and NSRF, Greece; ISF, MINERVA, GIF, I-CORE and Benoziyo Center,
Israel; INFN, Italy; MEXT and JSPS, Japan; CNRST, Morocco; FOM and NWO,
Netherlands; BRF and RCN, Norway; MNiSW and NCN, Poland; GRICES and FCT, Portugal; MNE/IFA, Romania; MES of Russia and ROSATOM, Russian Federation; JINR; MSTD,
Serbia; MSSR, Slovakia; ARRS and MIZ\v{S}, Slovenia; DST/NRF, South Africa;
MINECO, Spain; SRC and Wallenberg Foundation, Sweden; SER, SNSF and Cantons of
Bern and Geneva, Switzerland; NSC, Taiwan; TAEK, Turkey; STFC, the Royal
Society and Leverhulme Trust, United Kingdom; DOE and NSF, United States of
America.

The crucial computing support from all WLCG partners is acknowledged
gratefully, in particular from CERN and the ATLAS Tier-1 facilities at
TRIUMF (Canada), NDGF (Denmark, Norway, Sweden), CC-IN2P3 (France),
KIT/GridKA (Germany), INFN-CNAF (Italy), NL-T1 (Netherlands), PIC (Spain),
ASGC (Taiwan), RAL (UK) and BNL (USA) and in the Tier-2 facilities
worldwide.
%----------------------------------------------
%%\end{document}

\bibliographystyle{JHEP}
\bibliography{paperRun1}

%%%%%%%%%%%%%%%%%%%%%%%%%%%%%%%%%%%%%%%%%%%%%%%%%%%%%%%%%%%%%%%%%%%%%%%%%%%%%%%
% Auxiliary plots
%%%%%%%%%%%%%%%%%%%%%%%%%%%%%%%%%%%%%%%%%%%%%%%%%%%%%%%%%%%%%%%%%%%%%%%%%%%%%%%

%\appendix
%\input{AuxiliaryPlots}

\clearpage
% ATLAS Collaboration author list
% Data extracted on 26-Jan-2015 for paper reference HIGG-2013-30
%% \documentclass[11pt]{article}
%% \usepackage{a4wide}\begin{document}
\begin{flushleft}
{\Large The ATLAS Collaboration}

\bigskip

G.~Aad$^{\rm 85}$,
B.~Abbott$^{\rm 113}$,
J.~Abdallah$^{\rm 152}$,
S.~Abdel~Khalek$^{\rm 117}$,
O.~Abdinov$^{\rm 11}$,
R.~Aben$^{\rm 107}$,
B.~Abi$^{\rm 114}$,
M.~Abolins$^{\rm 90}$,
O.S.~AbouZeid$^{\rm 159}$,
H.~Abramowicz$^{\rm 154}$
% ...
\bigskip
\\
% ...
$^{11}$ Institute of Physics, Azerbaijan Academy of Sciences, Baku, Azerbaijan\\
% ...
$^{85}$ CPPM, Aix-Marseille Universit{\'e} and CNRS/IN2P3, Marseille, France\\
% ...
$^{90}$ Department of Physics and Astronomy, Michigan State University, East Lansing MI, United States of America\\
% ...
$^{107}$ Nikhef National Institute for Subatomic Physics and University of Amsterdam, Amsterdam, Netherlands\\
% ...
$^{113}$ Homer L. Dodge Department of Physics and Astronomy, University of Oklahoma, Norman OK, United States of America\\
$^{114}$ Department of Physics, Oklahoma State University, Stillwater OK, United States of America\\
% ...
$^{117}$ LAL, Universit{\'e} Paris-Sud and CNRS/IN2P3, Orsay, France\\
% ...
$^{152}$ Institute of Physics, Academia Sinica, Taipei, Taiwan\\
% ...
$^{154}$ Raymond and Beverly Sackler School of Physics and Astronomy, Tel Aviv University, Tel Aviv, Israel\\
% ...
$^{159}$ Department of Physics, University of Toronto, Toronto ON, Canada\\
% ...
\end{flushleft}

%% \end{document}
% Created with ./xml2latex.py

\end{document}